\documentclass[12pt]{article}
\usepackage{amsmath,amssymb,epsfig}
\makeatletter \@addtoreset{equation}{section} \makeatother
\renewcommand{\theequation}{\thesection.\arabic{equation}}
\newcommand{\ba}{\begin{array}}
\newcommand{\ea}{\end{array}}
\newcommand{\beq}{\begin{equation}}
\newcommand{\eeq}{\end{equation}}
\newcommand{\bea}{\begin{eqnarray}}
\newcommand{\eea}{\end{eqnarray}}
\def\bce{\begin{center}}
\def\ece{\end{center}}

\def\nonu{\nonumber}

\def\pa{\partial}

\def\be{\beta}

\def\de{\delta}

\def\ep{\epsilon}

\def\eps6{{\displaystyle \mathop{\epsilon}^{6}}{}}

\def\nab6{{\displaystyle \mathop{\nabla}^{6}}{}}
\def\0{{\sst{(0)}}}
\def\1{{\sst{(1)}}}
\def\2{{\sst{(2)}}}
\def\3{{\sst{(3)}}}
\def\4{{\sst{(4)}}}
\def\5{{\sst{(5)}}}
\def\6{{\sst{(6)}}}
\def\7{{\sst{(7)}}}
\def\8{{\sst{(8)}}}

\def\ba{\begin{array}}
\def\ea{\end{array}}
\def\beq{\begin{equation}}
\def\eeq{\end{equation}}
\def\be{\begin{equation}}
\def\ee{\end{equation}}

\def\Tr{\mathop{\rm Tr}}

\def\eps{\epsilon}

\def\ba{\begin{array}}
\def\ea{\end{array}}
\def\beq{\begin{equation}}
\def\eeq{\end{equation}}
\def\be{\begin{equation}}
\def\ee{\end{equation}}

\def\Tr{\mathop{\rm Tr}}

\def\eps{\epsilon}

\def\eps6{{\displaystyle \mathop{\epsilon}^{6}}{}}

\def\nab6{{\displaystyle \mathop{\nabla}^{6}}{}}

\newcommand{\HGF}[4]{{}_2 F_1 \left(#1,#2;#3;#4\right)}

\newcommand{\bean}{\begin{eqnarray*}}
\newcommand{\eean}{\end{eqnarray*}}

\begin{document}
\thispagestyle{empty} \addtocounter{page}{-1}
   \begin{flushright}
%KIAS-P08nnn \\
%CALT-68-nnnn \\
%{\tt hep-th/yymmnnn}\\
\end{flushright}

\vspace*{1.3cm}
  
\centerline{ \Large \bf  The Gauge Dual of  }
\vspace{.3cm} 
\centerline{ \Large \bf   A Warped Product of
  $AdS_4$ and }
\vspace{.3cm} 
\centerline{ \Large \bf  A Squashed and Stretched Seven-Manifold   } 
\vspace*{1.5cm}
\centerline{{\bf Changhyun Ahn  {\rm and} Kyungsung Woo }
%, {\bf Kazuo Hosomichi $^{2}$}
%and {\bf Sungjay Lee $^{2}$} 
} 
\vspace*{1.0cm} 
\centerline{\it  
Department of Physics, Kyungpook National University, Taegu
702-701, Korea} 
%\centerline{\it $^{2}$ Korea Institute for 
%Advanced Study, Seoul 130-012, Korea }
\vspace*{0.8cm} 
\centerline{\tt ahn@knu.ac.kr \qquad wooks@knu.ac.kr
} 
\vskip2cm

\centerline{\bf Abstract}
\vspace*{0.5cm}

Corrado, Pilch and Warner in 2001
have found the second 11-dimensional solution
where the deformed geometry of ${\bf S}^7$ in the lift contains ${\bf
  S}^2 \times {\bf S}^2$. 
We identify the gauge dual of this background with 
the theory described by Franco, Klebanov and Rodriguez-Gomez
recently. It is the $U(N) \times U(N) \times U(N)$ gauge theory 
with two $SU(2)$ doublet chiral fields $B_1$ transforming
in the   $({\bf N}, \overline{\bf N}, {\bf 1})$, 
 $B_2$ transforming
in the   $({\bf 1}, {\bf N}, \overline{\bf N})$,
 $C_1$ 
in the   $({\bf 1}, \overline{\bf N}, {\bf N})$
and 
 $C_2$ 
in the   $(\overline{\bf N}, {\bf N}, {\bf 1})$ as well as an adjoint
field $\Phi$ in the ({\bf 1}, {\bf adj}, {\bf 1}) representation.

By adding the mass term
for adjoint field $\Phi$,  
the detailed correspondence between fields of
$AdS_4$ supergravity and composite operators of the 
IR field theory is determined.
Moreover, we compute the spin-2 KK modes around 
a warped product of $AdS_4$ and a squashed and stretched seven-manifold. 
This background with global $SU(2) \times SU(2)  \times U(1)_R$ 
symmetry is related to a $U(N) \times U(N) \times U(N)$ 
${\cal N}=2$ superconformal Chern-Simons matter
theory 
with eighth-order superpotential and Chern-Simons level $(1,1,-2)$. 
The mass-squared in $AdS_4$ depends
on $SU(2) \times SU(2) \times U(1)_R$
quantum numbers and KK excitation number.  
The dimensions of spin-2 operators are found.

\baselineskip=18pt
\newpage
\renewcommand{\theequation}
{\arabic{section}\mbox{.}\arabic{equation}}

%%%%%%%%%%%%%%%%%%%%%%%%%%%%%%%%%%%%%%%%%%%%%%%%%%%%%%%%%%%%%%%%%%%%%
%%%%%%%%%%%%%%%%%%%%%%%%%%%%%%%%%%%%%%%%%%%%%%%%%%%%%%%%%%%%%%%%%%%%%%
\section{Introduction}
%%%%%%%%%%%%%%%%%%%%%%%%%%%%%%%%%%%%%%%%%%%%%%%%%%%%%%%%%%%%%%%%%%%%%%
%%%%%%%%%%%%%%%%%%%%%%%%%%%%%%%%%%%%%%%%%%%%%%%%%%%%%%%%%%%%%%%%%%%%%

The
${\cal N}=6$ $U(N) \times U(N)$ 
Chern-Simons matter theory
with level $k$ in 3-dimensions
is described as the low energy limit of $N$ M2-branes at 
${\bf C}^4/{\bf Z}_k$ singularity \cite{ABJM}.
For $k=1, 2$, the full ${\cal N}=8$ supersymmetry
is preserved while for $k > 2$,
the supersymmetry is broken to ${\cal N}=6$. 
The matter contents and the superpotential  of this theory
are the same as for the D3-branes on the conifold \cite{KW}. 
The RG flow
between the UV fixed point and 
the IR fixed point of the 3-dimensional 
gauge theory can be obtained from gauged ${\cal N}=8$ 
supergravity in 4-dimensions via AdS/CFT 
correspondence \cite{Maldacena}. 
The holographic
RG flow equation connecting ${\cal N}=8$ $SO(8)$ fixed point 
to ${\cal N}=2$ $SU(3) \times U(1)$ fixed point has been studied in 
\cite{AP,AW} where the $U(1)$ symmetry can be identified with $U(1)_R$
symmetry of 3-dimensional theory coming from the ${\cal N}=2$ supersymmetry while
those from ${\cal N}=8$ $SO(8)$ fixed point 
to
${\cal N}=1$ $G_2$
fixed point has been also studied in 
\cite{AW,AI,AR99}.
The M-theory lifts of these 
RG flows have been found in \cite{CPW,AI} by solving the
Einstein-Maxwell equations in 11-dimensions with the appropriate field
strengths in the internal space.

The mass deformed $U(2) \times U(2)$
Chern-Simons matter theory with level $k=1, 2$ 
preserving global $SU(3) \times U(1)_R$ symmetry has been studied 
in \cite{Ahn0806n2,BKKS,KKM,KPR} by identifying the turning on the
supergravity fields with the mass term in the boundary gauge theory while
the mass deformation for this theory preserving $G_2$
symmetry  has been described in \cite{Ahn0806n1}.
Due to the ${\cal N}=1$ supersymmetry for the latter, there is no
$U(1)_R$ symmetry.  
The nonsupersymmetric 
RG flow equations preserving $SO(7)^{\pm}$ symmetries 
have been discussed in \cite{Ahn0812} by looking at the domain wall solutions.  
The holographic
RG flow equations connecting ${\cal N}=1$ $G_2$ fixed point 
to ${\cal N}=2$ $SU(3) \times U(1)_R$ fixed point have been 
found in \cite{BHPW} by analyzing the mass terms for the scalar
potential at each critical point.  
Moreover, the ${\cal N}=4$ and ${\cal N}=8$ RG flows have been 
studied in \cite{AW09} by studying the explicit gauged supergravity theory. 
Recently, further developments on  
the gauged ${\cal N}=8$ supergravity in four-dimensions have been done
in \cite{AW09-1,Ahn0905} in the context of bulk and boundary theory. 
Very recently, by following the prescription of 
\cite{KPR}, the spin-2 Kaluza-Klein modes around 
a warped product of $AdS_4$ and a seven-ellipsoid which has 
global $G_2$ symmetry are discussed in \cite{AW0907} with the
computation of Laplacian eigenvalue problem. 

Are there any further examples which have an explicit $AdS_4/CFT_3$ 
correspondence?
The seven-sphere ${\bf S}^7$ can be realized by ${\bf S}^1$-fibration
over ${\bf CP}^3$ \cite{NP,Ahn0809}. For the standard Fubini-Study
metric on the ${\bf CP}^3$, it contains ${\bf CP}^2$
\cite{PW,CPW,AI02-1,Ahn0810} 
or ${\bf CP}^1
\times {\bf CP}^1$ \cite{CPW,CLP} inside of ${\bf CP}^3$.
It is natural to generalize the above seven-sphere to the $U(1)$
bundle over an arbitrary Einstein-Kahler manifold \cite{CPW}.
Although it is not known how to generalize the ${\bf CP}^3$ to an
arbitrary Einstein-Kahler 3-fold, an arbitrary Einstein-Kahler 2-fold can
replace the above ${\bf CP}^2$. The natural choice is given by the
above ${\bf CP}^1 \times {\bf CP}^1$. The $U(1)$ bundle over 
${\bf CP}^1 \times {\bf CP}^1$ is known as $T^{1,1}$ space.    
In \cite{CPW}, they have found two different 11-dimensional solutions
where the first contains ${\bf CP}^2$ with $SU(3) \times U(1)_R$
symmetry
and the second has ${\bf CP}^1
\times {\bf CP}^1$ with $SU(2) \times SU(2) \times U(1)_R$ symmetry. 
In this paper, we focus on the second solution. For the first
solution, there are many relevant works given 
in \cite{Ahn0806n2,BKKS,KKM,KPR}.
The structure of 11-dimensional metric is fixed by requiring 
that the maximally superymmetric $SO(8)$ vacuum should preserve the 
$AdS_4 \times {\bf S}^7$ solution. This leads to the above $T^{1,1}$
space for particular coordinate inside the 11-dimensional metric. 
Furthermore, they have found that the Ricci tensor for the second
solution with frame basis is exactly the same as the one of the first 
solution by assuming that the supergravity fields satisfy the same
equations of motion discovered by \cite{AP}. For the 3-form potential 
the two solutions are different from each other. 
It is surprising that the same flow equations in 4-dimensions provide
two different 11-dimensional solutions to the equations of the
11-dimensional supergravity theory.

Then it is natural to ask what is the dual gauge theory corresponding
to the above second 11-dimensional solution.
Recently,  Franco, Klebanov and Rodriguez-Gomez \cite{FKR} have
studied the M2-branes on resolved cones over $Q^{1,1,1}$, denoted by 
${\cal C}(Q^{1,1,1})$, motivated by 
the observations for D3-branes in \cite{KW99} in type IIB theory. 
Since ${\cal C}(Q^{1,1,1})$ can be described by 
${\bf C}^2$ bundle over ${\bf CP}^1 \times {\bf CP}^1$, blowing-up one
${\bf CP}^1$ leads to ${\cal C}(T^{1,1}) \times {\bf C}$ where 
${\cal C}(T^{1,1})$ is cone over $T^{1,1}$ space.  
The blowing-up one
${\bf CP}^1$ corresponds to removing a point in the toric diagram.
The resolutions correspond to turning on the vevs for the scalar
component of chiral superfield.
Among two possible blow-ups, one of them provides the dual gauge
theory we describe in this paper. 
Originally, the quiver Chern-Simons gauge theory dual to $AdS_4 \times
Q^{1,1,1}$ is characterized by $U(N)_1 \times U(N)_2 \times U(N)_3 \times
U(N)_4$ gauge theory with levels $(1,1,-1,-1)$ coupled to three kinds of
bifundamental chiral superfields with a sextic superpotential 
\cite{FHPR}. The symmetry here is given by 
$SU(2) \times U(1) \times U(1)_R$ which is smaller than 
$SU(2) \times SU(2) \times SU(2) \times U(1)_R$ that is  
the symmetry of $Q^{1,1,1}$ space itself.
By turning on the vev for one of the chiral superfields and renaming
the remaining doublet to the adjoint field, the superpotential reduces
to the interaction between this adjoint field and other two kinds of
chiral superfields. Then the groups $U(N)_3 \times U(N)_4$ breaks into
the diagonal $U(N)$ subgroup leading to three product gauge groups. 
The previous Chern-Simons level becomes $(1,1,
-2)$.    

The
supersymmetric flow solution \cite{AP,AW} in four dimensional ${\cal N}=8$ gauged
supergravity
interpolates between an exterior $AdS_4$ region with maximal
supersymmetry
and an interior $AdS_4$ with one quarter of the maximal supersymmetry.  
This can be interpreted as the RG flow in ${\cal N}=8$ theory
which has $OSp(8|4)$ symmetry
broken to an ${\cal N}=2$ theory which has $OSp(2|4)$ symmetry
by the addition of a mass term for the adjoint chiral superfield.
The role of this massive adjoint superfield is completely different
from those in $SU(3) \times U(1)_R$ symmetric case.
A precise correspondence 
is obtained between fields of bulk supergravity in the $AdS_4$ region
and composite operators of the IR field theory in three dimensions. 
The global symmetry can be obtained from the structure of ${\bf CP}^1
\times {\bf CP}^1$ whose symmetry group is $SU(2) \times SU(2)$ and
the ${\cal N}=2$ supersymmetry which gives $SO(2)_R=U(1)_R$ symmetry.
We calculate the explicit KK spectrum of the spin-2 fields in 
$AdS_4$ by following the  work of \cite{KPR,AW0907}.
In an 11-dimensional theory, the equation for the metric
perturbations leads to a minimally coupled scalar equation. 
We obtain all the KK modes in terms of the seven variables
parametrizing the deformed seven-manifold.
The squared-mass terms in $AdS_4$ for all the modes depend on
the $SU(2) \times SU(2) \times U(1)_R$ quantum number and the 
KK excitation number.
We describe the corresponding ${\cal N}=2$ dual SCFT operators.

In section 2, the 11-dimensional background \cite{CPW} is reviewed. 
In section 3,  the $OSp(2|4)$ representations and $SU(2) \times SU(2)
\times U(1)_R$ representations in the supergravity mass spectrum for
each multiplet at the ${\cal N}=2$ critical point and the
corresponding ${\cal N}=2$ superfields in the dual gauge theory are
given in the spirit of \cite{Ahn0806n2,Ahn0806n1}. 
The Kahler potential is also obtained.
In section 4, the minimally coupled scalar equation \cite{KPR,AW0907} 
in the background
of section 2 and the mass formula are determined. The quantum numbers of
the ${\cal N}=2$ SCFT operators 
in Chern-Simons matter
theory with eighth-order superpotential are given.     
In section 5, the summary of this paper is given and the future
directions also are given.
In the Appendices A, B and C, 
we present some details for the field strength and
Ricci tensor.

%%%%%%%%%%%%%%%%%%%%%%%%%%%%%%%%%%%%%%%%%%%%%%%%%%%%%%%%%%%%%%%%
%%%%%%%%%%%%%%%%%%%%%%%%%%%%%%%%%%%%%%%%%%%%%%%%%%%%%%%%%%%%%%%%
\section{An ${\cal N}=2$ supersymmetric 
$SU(2) \times SU(2)  \times U(1)_R$-invariant 
flow in an 11-dimensional theory}
%%%%%%%%%%%%%%%%%%%%%%%%%%%%%%%%%%%%%%%%%%%%%%%%%%%%%%%%%%%%%%%%
%%%%%%%%%%%%%%%%%%%%%%%%%%%%%%%%%%%%%%%%%%%%%%%%%%%%%%%%%%%%%%%%

Let us review the 11-dimensional uplift of the supergravity background 
with global $SU(2) \times SU(2) \times U(1)_R$ 
symmetry and recall that  the supergravity background 
with global $SU(3) \times U(1)_R$ 
was found in \cite{CPW} as a nontrivial extremum
of the gauged ${\cal N}=8$ supergravity in 4-dimensions.  
The 11-dimensional coordinates
with indices $M, N, \cdots$ are decomposed into 4-dimensional spacetime 
$x^{\mu}$ and
7-dimensional internal space $y^m$.
Denoting the 11-dimensional metric as $g_{MN}$ with 
the convention 
$(-, +, \cdots, +)$
and the antisymmetric 
tensor fields as $F_{MNPQ} = 4\,\pa_{[M} A_{NPQ]}$, the
Einstein-Maxwell equations are given by \cite{CJS}
\bea
R_{M}^{\;\;\;N} & = & \frac{1}{3} \,F_{MPQR} F^{NPQR}
-\frac{1}{36} \de^{N}_{M} \,F_{PQRS} F^{PQRS},
\nonu \\
\nabla_M F^{MNPQ} & = & -\frac{1}{576} \,E \,\ep^{NPQRSTUVWXY}
F_{RSTU} F_{VWXY},
\label{fieldequations}
\eea
where the covariant derivative $\nabla_M$ 
on $F^{MNPQ}$ in 
(\ref{fieldequations})
is given by 
$E^{-1} \pa_M ( E F^{MNPQ} )$ together with elfbein determinant 
$E \equiv \sqrt{-g_{11}}$. The epsilon tensor 
 $\ep_{NPQRSTUVWXY}$ with lower indices is purely numerical.
The geometry is a warped product of $AdS_4$ and the
squashed and stretched 7-dimensional manifold.

The conifold \cite{Cd} that is a singular noncompact Calabi-Yau three-fold
is a surface in ${\bf C}^4$ parametrized by four complex 
coordinates $z_1, z_2, z_3$ and $z_4$ which
satisfy the quadratic equation 
\bea
z_1^2 + z_2^2 + z_3^2 +z_4^2 =0,
\label{cond}
\eea
where these are given by six angular variables 
\bea
z_1 & = & \frac{r^{\frac{3}{2}}}{\sqrt{2}} \left[ \cos
  (\frac{\theta_1}{2}) \cos (\frac{\theta_2}{2}) \, e^{\frac{i}{2} (\psi
   +\phi + \phi_1 + \phi_2)} - \sin (\frac{\theta_1}{2}) \sin
  (\frac{\theta_2}{2}) \, e^{\frac{i}{2} (\psi+\phi-\phi_1-\phi_2)} \right],
\nonu \\
z_2 & = & \frac{r^{\frac{3}{2}}}{\sqrt{2} i} \left[ -\cos
  (\frac{\theta_1}{2}) \cos (\frac{\theta_2}{2}) \, e^{\frac{i}{2} (\psi
    +\phi + \phi_1 + \phi_2)} - \sin (\frac{\theta_1}{2}) \sin
  (\frac{\theta_2}{2}) \, e^{\frac{i}{2} (\psi+\phi-\phi_1-\phi_2)} \right],
\nonu \\
z_3 & = & \frac{r^{\frac{3}{2}}}{\sqrt{2}} \left[ -\cos
  (\frac{\theta_1}{2}) \sin (\frac{\theta_2}{2}) \, e^{\frac{i}{2} (\psi
    + \phi + \phi_1 - \phi_2)} - \sin (\frac{\theta_1}{2}) \cos
  (\frac{\theta_2}{2}) \, e^{\frac{i}{2} (\psi+\phi-\phi_1+\phi_2)} \right],
\nonu \\
z_4 & = & \frac{r^{\frac{3}{2}}}{\sqrt{2} i} \left[ -\cos
  (\frac{\theta_1}{2}) \sin (\frac{\theta_2}{2}) \, e^{\frac{i}{2} (\psi
    +\phi + \phi_1 - \phi_2)} + \sin (\frac{\theta_1}{2}) \cos
  (\frac{\theta_2}{2}) \, e^{\frac{i}{2} (\psi+\phi-\phi_1+\phi_2)} \right].
\label{zis}
\eea
Here one $SU(2)$ acts on $\theta_1, \phi_1$ and $\phi$ while the other
acts on $\theta_2, \phi_2$ and $\psi$. See also earlier work by \cite{Romans}.
One can easily check that
$\sum_{i=1}^{4}|z_i|^2 = r^3$.
In order to introduce 7-dimensional coordinatization, one multiplies 
the above four complex coordinates by $\cos \mu$ as follows: 
\bea
z_1 \rightarrow \cos\mu \,z_1, \qquad
 z_2 \rightarrow \cos\mu \,z_2, \qquad
z_3 \rightarrow \cos\mu \,z_3, \qquad
z_4 \rightarrow \cos\mu \,z_4, 
\label{shift}
\eea
and moreover, we introduce the other complex coordinate
\bea
w = \sin\mu \, r^{\frac{3}{2}} \, e^{-i\psi},
\label{w}
\eea
with the property
$
1- |w|^2 = r^3 \cos^2 \mu$.
Then one obtains
$
\sum_{i=1}^{4} |z_i|^2 +|w|^2 = r^3$,
and for $r=1$, this leads to a seven-sphere of radius $1$ in ${\bf
  R}^8$ due to the constraint (\ref{cond}).

How do we obtain the second 11-dimensional solution? One needs to 
find out the right metric structure first. 
By replacing the ${\bf CP}^2$ inside the 11-dimensional metric 
with ${\bf CP}^1 \times {\bf CP}^1$, 
the second 
eleven-dimensional metric in \cite{CPW} preserving $SU(2) \times
SU(2) \times U(1)_R$ where the ${\cal N}=2$ supersymmetry is observed
via $U(1)_R$ symmetry in 3-dimensions is, by evaluating the nontrivial
warp factor coming from 4-dimensional supergravity data, given by 
\bea
ds_{11}^2 =\Delta^{-1}\left(dr^2 +e^{2 A(r)}
\eta_{\mu\nu}dx^\mu dx^\nu \right)+3^{\frac{3}{2}} \hat{L}^2
\sqrt{\Delta} \,
ds_7^2(\rho,\chi),
\label{11dmetric}
\eea
where $\eta_{\mu \nu}=(-,+,+)$ and 
the 7-dimensional metric in terms of supergravity fields $\rho$
and $\chi$ is
\bea
ds_7^2(\rho,\chi) & = & \frac{X}{\rho^6} \, d \mu^2 +
\rho^2 \cos^2 \mu \left[ \frac{1}{6} (d \theta_1^2 + \sin^2\theta_1 \,
  d\phi_1^2) + \frac{1}{6} (d \theta_2^2 + \sin^2\theta_2 \, d
  \phi_2^2) \right]
\nonu \\
& + & \frac{\rho^{10}}{4X} \sin^2 2\mu \left[ - \frac{d \psi}{\rho^8}
  + \frac{1}{3}(d \psi + d \phi + \cos \theta_1 \, d \phi_1 + 
\cos \theta_2 \, d \phi_2)
 \right]^2
\nonu \\
& + & \frac{\rho^2 \cosh^2 \chi}{X} \left[ \sin^2\mu \, d \psi + 
\frac{1}{3} \cos^2 \mu (d \psi + d \phi + \cos \theta_1 \, d \phi_1 + 
\cos \theta_2 \, d \phi_2)\right]^2,
\label{7dmet2}
\eea
and we introduce 
\bea
X \equiv \cos^2 \mu + \rho^8 \sin^2 \mu,
\label{X}
\eea
and 
\bea
\Delta = \frac{\rho^{\frac{4}{3}}}{ X^{\frac{2}{3}} \, \cosh^{\frac{4}{3}} \chi }.
\label{delta}
\eea
The $A(r)$ in (\ref{11dmetric}) is a scale factor.
The main results of this background are the fact that they showed the equations
of motion for $\rho$ and $\chi$ are the same as the one in $SU(3)
\times U(1)_R$ invariant RG flows and the corresponding 3-form 
has little modification. 
In particular, for the $\mu=0$, the above (\ref{7dmet2}) becomes 
\bea
\rho^2 ds^2_{T^{1,1}} +\rho^2 \sinh^2 \chi \,
\frac{1}{9} (d \psi  + d \phi + \cos \theta_1 \, d \phi_1 + 
\cos \theta_2 \, d \phi_2)^2,
\label{modu}
\eea
where the five-dimensional metric is well-known $T^{1,1}$ manifold 
\bea
ds^2_{T^{1,1}}  & = & \frac{1}{9} (d \psi  + d \phi + \cos \theta_1 \, d \phi_1 + 
\cos \theta_2 \, d \phi_2)^2 \nonu \\
&+& 
\frac{1}{6} (d \theta_1^2 + \sin^2\theta_1
 \, d\phi_1^2) + \frac{1}{6} (d \theta_2^2 + \sin^2\theta_2  \, d
  \phi_2^2).
\label{t11}
\eea
For large $r$($\rho \rightarrow 1$ and $\chi \rightarrow 0$), the
moduli space (\ref{modu}) approaches the Ricci flat Kahler conifold.  
We are focusing on the IR critical values in the background of (\ref{11dmetric})
\bea
\rho = 3^{\frac{1}{8}}, \qquad \chi =\frac{1}{2} \cosh^{-1} 2.
\label{rhochi}
\eea
The amounts of squashing and stretching are parametrized by $\rho$ and
$\chi$
respectively.

Secondly, we need to find out the right 3-form structure.
One can start with the 3-form on the 7-dimensional space
\bea
C^{(3)} = -\frac{1}{4} \sinh \chi \, e^{i(-\phi-2\psi)} \, (e^5+ i e^{10}) 
\wedge (e^6 + i e^7) \wedge (e^8 + i e^9),
\label{c3}
\eea
in order to describe the 4-form $F^{(4)}$ that solves
(\ref{fieldequations})
in the frame basis
together 
with (\ref{11dmetric}) \footnote{
Let us emphasize that the form of (\ref{c3}) is different from the one
in \cite{CPW} where maybe the careless typing was done. 1) The imaginary
$i$ should be in the exponent of exponential function  and 2) the
coefficient in $e^{10}$ should be changed in the original
paper \cite{CPW}.}.

Then we take 3-form in 11-dimensional geometry as
\bea
A^{(3)} = \frac{3^{\frac{3}{4}}}{4} e^{\frac{3r}{\hat{L}}} \, d x^1
\wedge d x^2 \wedge d x^3 + C^{(3)} + (C^{(3)})^{\ast}.
\label{a3}
\eea
The internal part of $F^{(4)}$ can be written as
$ d  C^{(3)} + d (C^{(3)})^{\ast}$.
The antisymmetric 
tensor fields can be obtained  from 
$F^{(4)} = d A^{(3)}$ with (\ref{a3}).
We present both the 4-form and the 
Ricci tensor in frame basis in the Appendix A.

Let us define the R-charge to be given by the Killing vector
\bea
R = -i \left(   2 \pa_{\phi} - \pa_{\psi} \right) =
\frac{1}{2} \left( z_i \pa_{z_i} -\overline{z}_i \pa_{\overline{z}_i} 
\right) + w \pa_{w}-\overline{w} \pa_{\overline{w}}.
\label{R}
\eea
Then the $z_i$'s coordinates have R-charge $\frac{1}{2}$ while the 
$w$ coordinate has R-charge $1$, from (\ref{zis}) and (\ref{w}). 
Note that the corresponding shifts in $\psi$ and $\phi$ under 
(\ref{R}), that is, $\phi \rightarrow \phi + 2 \gamma$ and $\psi
\rightarrow \psi - \gamma$ preserve the quantity $\phi +2 \psi$.
Of course, this $U(1)$ should be identified with the $U(1)_R$ symmetry 
of the dual gauge theory we will discuss next sections.
Therefore, the ambiguity for the $U(1)_R$ charge 
in group theory analysis alone, 
coming from the different embeddings of
$U(1)_R$ in $SO(8)$, is resolved by the above 
$U(1)_R$ charge assignment (\ref{R}).  

%%%%%%%%%%%%%%%%%%%%%%%%%%%%%%%%%%%%%%%%%%%%%%%%%%%%%%%%%%%%%%%%
%%%%%%%%%%%%%%%%%%%%%%%%%%%%%%%%%%%%%%%%%%%%%%%%%%%%%%%%%%%%%%%%
\section{$OSp(2|4)$ spectrum and operator map between bulk and
  boundary theories}
%%%%%%%%%%%%%%%%%%%%%%%%%%%%%%%%%%%%%%%%%%%%%%%%%%%%%%%%%%%%%%%%
%%%%%%%%%%%%%%%%%%%%%%%%%%%%%%%%%%%%%%%%%%%%%%%%%%%%%%%%%%%%%%%%

What is the dual gauge theory corresponding to the previous 
11-dimensional background in the context of AdS/CFT?
By giving a vacuum expectation value to one of the internal fields in
the quiver $U(N)\times U(N) \times U(N) \times U(N)$ 
Chern-Simons gauge theory for M2-branes probing 
the cone over $Q^{1,1,1}$, ${\cal C}(Q^{1,1,1})$, the theory becomes 
the quiver diagram for a partial resolution of $Q^{1,1,1}$ theory
with the gauge group \cite{FKR}
$
U(N) \times U(N) \times U(N)$,
and the $SU(2)$ doublets chiral fields are given by
\bea
&& B_1 \;\;\; \mbox{in} \;\;\; ({\bf N}, \overline{\bf N}, {\bf 1}), 
\qquad B_2 \;\;\; \mbox{in} \;\;\; ({\bf 1}, {\bf N},
\overline{\bf N}), \nonu \\
&& C_1 \;\;\; \mbox{in} \;\;\; ({\bf 1}, \overline{\bf N}, {\bf N}), 
\qquad C_2 \;\;\; \mbox{in} \;\;\; (\overline{\bf N}, {\bf
  N}, {\bf 1}),
\label{b1b2c1c2}
\eea
and there exists an adjoint field 
\bea
\Phi \;\;\; \mbox{in} \;\;\; ({\bf 1}, {\bf adj}, {\bf 1}).
\label{Phi}
\eea
The superpotential is given by the following interaction between
these fields 
\bea
W = \Phi \left( C_2 B_1 B_2 C_1 - B_2 C_1 C_2 B_1 \right).
\label{W}
\eea
What are the scale dimensions for these fields?
The scale dimension of four chiral superfields are $\frac{3}{8}$ which
is equal to $\frac{1}{2} \times \frac{3}{4}$ at
the UV(the number $\frac{1}{2}$ is checked in \cite{Ahn0806n2}) 
while the one of adjoint superfield becomes $\frac{1}{2}$ in
order to have vanishing beta-function. The ratio $\frac{3}{4}$ is
equivalent to the ratio of the number of massless fields in $SU(3)
\times U(1)_R$ \cite{Ahn0806n2,BKKS} 
to the number of massless fields in $SU(2) \times SU(2)
\times U(1)_R$ here: (\ref{Phi}).  
Then the scale dimension of (\ref{W}) is two. 
See also \cite{LLP} where they computed the $U(1)_R$ charges for
${\cal C}(T^{1,1}) \times {\bf C}$ Calabi-Yau 4-fold as $(\frac{3}{8},
\frac{3}{8}, \frac{3}{8}, \frac{3}{8}, \frac{1}{2})$.
The Chern-Simons level $\vec{k}=(1,1,-1,-1)$ of $Q^{1,1,1}$ is transformed as
$(1,1,-2)$.
The invariants are given by \cite{FKR}
\bea
z_1 = B_1 C_2, \qquad z_2 = B_2 C_1, \qquad z_3 = B_1 C_1, \qquad z_4
= B_2 C_2, \qquad w = \Phi,
\label{dualzis}
\eea
where $z_i(i=1, 2, 3, 4)$ parametrize the conifold via (\ref{zis}) 
and the adjoint
field parametrizes the complex line ${\bf C}$.
Due to the complex field $\Phi$ which has a superpotential giving it a
mass which drives the flow, we introduce the following mass term 
\bea
\Delta W = \frac{1}{2} m  \Phi^2,
\label{quad}
\eea
where the IR value of scaling dimension for $\Phi$ is $1$
with scale dimension zero for the mass $m$.
By integrating out the massive field in the superpotential
$W+\Delta W$ given by (\ref{W}) and (\ref{quad}), one obtains the effective new 
eighth-order superpotential 
\bea
W_{eff} = \frac{1}{m}  \left( C_2 B_1 B_2 C_1 - B_2 C_1 C_2 B_1 \right)^2.
\label{weff}
\eea
How do we check the scale dimensions?
This implies that the IR value of scaling dimension for $B_i$ and
$C_i$ from (\ref{weff}) is 
$\frac{1}{4}$ which is identical to $\frac{1}{3} \times \frac{3}{4}$. 
Again, the number $\frac{1}{3}$ is checked in \cite{Ahn0806n2} and 
the origin for the number $\frac{3}{4}$ is explained before: the
ratio of the number of masssless fields in two different cases.  
See also \cite{JLP01}.

From the branching rule \cite{ps} of $SO(8)$ into 
$SU(2) \times SU(2)$, the spin $2, \frac{3}{2},
1, \frac{1}{2}, 0$ fields transform as 
\bea
{\bf 1} & \rightarrow & ({\bf 1},{\bf 1}),
\nonu \\
{\bf 8} & \rightarrow & 4 ({\bf 1},{\bf 1})   
\oplus ({\bf 2},{\bf 2}),
\nonu \\
{\bf 28} & \rightarrow & 
 6 ({\bf 1},{\bf 1})   
\oplus 4 ({\bf 2},{\bf 2}) \oplus  ({\bf 1},{\bf 3}) \oplus 
({\bf 3},{\bf
1}),
\nonu \\
{\bf 56}  & \rightarrow &    4 ({\bf 1},{\bf 1})   
\oplus 7 ({\bf 2},{\bf 2}) \oplus 4 ({\bf 1},{\bf 3}) \oplus 4
({\bf 3},{\bf
1}),
\nonu \\
{\bf 70}  & \rightarrow & 
 11 ({\bf 1},{\bf 1})   
\oplus  8 ({\bf 2},{\bf 2}) \oplus  3 ({\bf 1},{\bf 3}) \oplus 3
({\bf 3},{\bf
1}) \oplus ({\bf 3},{\bf 3}),
\label{decom}
\eea
respectively. Fields of different spin but same $SU(2) \times SU(2)$
representation in the decomposition (\ref{decom}) 
of the ${\cal N}=8$ supermultiplet
must recombine into various ${\cal N}=2$ supermultiplets.
The correspondence between fields of
$AdS_4$ supergravity and composite operators of the 
IR field theory can be described in this section, along the spirit of 
\cite{Ahn0806n1,Ahn0806n2}.

The even subalgebra of the superalgebra $OSp(2|4)$
is a direct sum of subalgebras
where $Sp(4,R)\simeq SO(3,2)$ is the isometry algebra of $AdS_4$
and the compact subalgebra $SO(2)$ generates $U(1)_R$ symmetry
\cite{Merlatti,CDDF1,FFGT1}.
The maximally compact subalgebra is then 
$SO(2)_E \times SO(3)_S \times SO(2)_Y$
where the generator of $SO(2)_E$ is the hamiltonian of the system
and its eigenvalues $E$ are the energy levels of states for the
system,
the group $SO(3)_S$ is the rotation group and its representation $s$ 
describes the spin states of the system, and 
the eigenvalue $y$ of the generator of $SO(2)_Y$ is the hypercharge of
the state.
A supermultiplet, a unitary irreducible representations(UIR) of the
superalgebra
$OSp(2|4)$, consists of a finite number of UIR of the even subalgebra
and a particle state is characterized by a spin $s$, a mass $m$ and a 
hypercharge $y$. The relations between the mass and energy are given
in \cite{CFN}.

Let us classify the supergravity multiplet which is invariant under 
$SU(2) \times SU(2) \times U(1)_Y$ and describe them in the three dimensional 
boundary theory.

%%%%%%%%%%%%%%%%%%%%%%%%%%%%%%%%%%%%%%%%
%%%%%%%%%%%%%%%%%%%%%%%%%%%%%%%%%%%%%%%%%
$\bullet$ Long massive vector multiplet
%%%%%%%%%%%%%%%%%%%%%%%%%%%%%%%%%%%%%%%%%
%%%%%%%%%%%%%%%%%%%%%%%%%%%%%%%%%%%%%%%%%

The conformal dimension $\Delta$
is given by $\Delta=E_0$ and  
the $U(1)_R$ charge is $0$.
The $U(1)_R$ charge 
is related to a hypercharge by
$
R = y$. 
The $K$ is a general unconstrained
scalar superfield in the boundary theory and 
has a dimension $ \frac{1}{2}(5+\sqrt{17})$ in the IR \cite{Ahn0806n2}
because we use the same 4-dimensional flow equations.

Let us describe the Kahler potential more detail and 
it
is found in \cite{JLP01}, by looking at the 11 dimensional flow
equation \cite{CPW},  as 
\bea
K = \frac{1}{4} \tau_{M2} L^2 e^A \left( \rho^2 + \frac{1}{\rho^6}
\right), \qquad \frac{d q}{d r} = \frac{2}{L \rho^2} q,
\label{sol}
\eea
where $\rho \equiv e^{\frac{\lambda}{4\sqrt{2}}}$ and $\chi \equiv 
\frac{\lambda'}{\sqrt{2}}$ in previous notations \cite{AP}.
The corresponding Kahler metric is given by \cite{JLP01}
\bea
d s^2 = \frac{1}{4p^2} \left( p \frac{d}{d p}\right)^2 K dp^2 +
\left( p \frac{d}{d p} \right) K d \hat{z}_i d \hat{\overline{z}}_i +
\left( p^2 \frac{d^2}{ d p^2} \right) K |\hat{z}_i  d \hat{\overline{z}}_i |^2,
\label{metric}
\eea
where the coordinate $p$ is defined as
\bea
p   \equiv  z_1 \overline{z}_1 + z_2 \overline{z}_2 + z_3
\overline{z}_3 + z_4 \overline{z}_4,
\label{p}
\eea
and 
the four complex coordinates parametrize the ${\bf C}^4$
and the $\hat{z}_i$'s are coordinates on an ${\bf S}^7$ of unit radius. 
The quantity $q$ will be defined later.
The $\frac{2}{3} d \hat{z}_i d
\hat{\overline{z}}_i-\frac{2}{9}|\hat{z}_i 
d \hat{\overline{z}}_i|^2$ is a metric on the $T^{1,1}$ (\ref{t11}) while 
$|\hat{z}_i d \hat{\overline{z}}_i|^2$ is the $U(1)$ fiber in the
description of $T^{1,1}$. 
Note that there is a relation given by the first order differential
equation
$\frac{d K}{d r} = \tau_{M_2} 
L e^A$ \cite{JLP01}.
The moduli space is parametrized by the vacuum expectation values of
the four massless scalars $\Phi_1, \Phi_2, \Phi_3$ and $\Phi_4$ denoted as
$z_1, z_2, z_3$ and $z_4$ from (\ref{dualzis})
\bea
\Phi_1 =z_1, \qquad 
\Phi_2 =z_2, \qquad
\Phi_3 =z_3, \qquad
\Phi_4 =z_4.
\label{Phi}
\eea
The $z_i(i=1, 2, 3, 4)$ transform in the
representation $({\bf 2}, {\bf 2})$ of 
$SU(2) \times SU(2)$ while their complex
conjugates $\overline{z}_i$ also transform in the representation 
$({\bf 2},{\bf 2})$.  

At the UV end of the flow which is just $AdS_4 \times {\bf S}^7$, 
$A(r) \sim \frac{2}{L} r$
from the solution 
for $A(r)$ and $W=1$ \cite{AP}. Moreover,
the radial coordinate on moduli space $\sqrt{q} \sim
e^{\frac{r}{ L}} \sim e^{\frac{A(r)}{2}}
$ 
from (\ref{sol}) by
substituting $\rho =1$. Therefore, the
Kahler potential from (\ref{sol}) behaves as $K \sim e^{A(r)} \sim q
\sim p^{\frac{2}{3}}$. Note that $q$ and $p$ has the following
relation $q  \equiv \frac{3}{2} p^{\frac{2}{3}}$ \cite{JLP01}. 
This implies that  $K = \left(\Phi_1 \overline{\Phi}_1 +\Phi_2 \overline{\Phi}_2+
\Phi_3 \overline{\Phi}_3 +\Phi_4 \overline{\Phi}_4
\right)^{\frac{2}{3}} $ 
at the UV in the boundary theory.
The scaling dimensions for $\Phi_i(i=1,2,3,4)$ and its conjugate
fields
are $\frac{3}{4}$ as explained around (\ref{W}) and (\ref{dualzis}). See also 
\cite{LLP} where the R-charges in a Calabi-Yau
four-fold,  the cone over $T^{1,1}$ multiplied by a complex line ${\bf
C}$(denoted by ${\cal C}(T^{1,1}) \times
{\bf C}$ as in the introduction), in the context of toric diagram are computed. 
Note that the scaling dimension of $K$ is equal to $1$
which is correct because it should have scaling dimension $1$.
The observation for this particular number  
$\frac{3}{4}$ is also noticed in \cite{JLP01} earlier.

At the IR end of the flow, $A(r) \sim \frac{3^{\frac{3}{4}}}{L} r$
with $SO(8)$ coupling $g \equiv \frac{\sqrt{2}}{L}$ from the solution 
for $A(r)$ and $W= \frac{3^{\frac{3}{4}}}{2}$ \cite{AP}. Moreover,
$\sqrt{q} \sim
e^{\frac{3^{-\frac{1}{4}} r}{ L}} \sim e^{\frac{A(r)}{3}}
$ 
from (\ref{sol}) by
substituting $\rho =3^{\frac{1}{8}}$ (\ref{rhochi}). Therefore, the
Kahler potential behaves as $K \sim e^{A(r)} \sim q^{\frac{3}{2}} \sim
p$. 
Then $K$ becomes from (\ref{p}) and (\ref{Phi})
\bea
K =\Phi_1 \overline{\Phi}_1 +\Phi_2 \overline{\Phi}_2+
\Phi_3 \overline{\Phi}_3  + \Phi_4 \overline{\Phi}_4,
\label{K}
\eea
in the boundary theory.
Obviously, from the tensor product between $({\bf 2},{\bf 2})$ and $({\bf
  2},{\bf 2})$
of $SU(2) \times SU(2)$ representation, one gets a singlet $({\bf
  1},{\bf 1})_0$ with
$U(1)_R$ charge $0$, as in Table 1. 
Note that $\Phi_i(i=1,2,3,4)$ has $U(1)_R$ charge 
$\frac{1}{2}$(explained around (\ref{weff})) \cite{JLP01} 
while $\overline{\Phi}_i(i=1,2,3,4)$ has $U(1)_R$ charge 
$-\frac{1}{2}$.
As observed in \cite{JLP01}, the number $\frac{1}{2}$ is equal to
simply $\frac{3}{4}$ times the dimension of fields in $SU(3) \times
U(1)_R$ flow. That is, $\frac{3}{4} \times \frac{1}{3}$ which should
be
multiplied by  $2$ because $\Phi_i$ is quadratic in the $SU(2)$
doublet chiral fields:(\ref{dualzis}) and (\ref{Phi}). 
Since the scaling dimensions for 
 $\Phi_i(i=1,2,3,4)$ and its conjugate
fields
are $\frac{1}{2}$, the scaling dimension of $K$ is $1$ which is
consistent with classical value as before.
The corresponding Kahler metric (\ref{metric}) 
provides the Kahler term in the action.
For the superfield $K$ (\ref{K}), the 
action looks like $\int d^3 x d^2 \theta^{+} d^2 \theta^{-} 
K$ as in \cite{Ahn0806n2}. 
The component content of this action 
can be worked out straightforwardly using the projection technique. 
This implies that the highest component field in $\theta^{\pm}$-expansion,
the last element in Table 1, has a conformal dimension  $\frac{1}{2} 
(5+\sqrt{17})$ in the IR as before. 

The corresponding $OSp(2|4)$ representations and corresponding
${\cal N}=2$ superfield in three dimensions are listed in Table 1.
The relation between $\Delta$ and the mass for various fields can be
found in \cite{CFN}. For spin $0$ and $1$, their relations are 
given by $\Delta_{\pm} = 
\frac{3 \pm \sqrt{1+\frac{m^2}{4}}}{2}$ where we have to choose the
correct root among two cases as in \cite{Ahn99} while for spin $\frac{1}{2}$,
the explicit form is given by $\Delta =\frac{6 + |m|}{4}$. 
Using these relations, one can read off the mass for each state.

%%%%%%%%%%%%%%%%%%%%%%%%%%%%%%%%%%%%%%%%%%%%%%%%%%%%%%%%%%%%%%%%%%%%%%
%table 1%%%%%%%%%%%%%%%%%%%%%%%%%%%%%%%%%%%%%%%%%%%%%%%%%%%%%%%%%%%%%%%%%%%%%
\begin{table} 
\begin{center}
\begin{tabular}{|c|c|c|c|c|} \hline
Boundary Operator(B.O.) & Energy & Spin $0$  & Spin $\frac{1}{2}$ & Spin $1$
 \\ \hline
$K =  \Phi_1 \overline{\Phi}_1 +\Phi_2 \overline{\Phi}_2 +
\Phi_3 \overline{\Phi}_3
 +\Phi_4 \overline{\Phi}_4 $ & $E_0=\frac{1}{2}(1+\sqrt{17})$ & $({\bf
 1},{\bf 1})_0$ &  &   \\
& $E_0+\frac{1}{2}
=\frac{1}{2}(2+\sqrt{17})
$ & & 
$({\bf 1},{\bf 1})_{\pm\frac{1}{2}}$  &
   \\
& $E_0+1
=\frac{1}{2}(3+\sqrt{17})
$ &  $({\bf 1},{\bf 1})_{\pm 1, 0} $ &  
& $({\bf 1},{\bf 1})_0$  \\
& $E_0+\frac{3}{2}
=\frac{1}{2}(4+\sqrt{17})
$ &  & $({\bf 1},{\bf 1})_{\pm\frac{1}{2}} $ &   \\
& $E_0+2 
= \frac{1}{2}(5+\sqrt{17})
$ & $({\bf 1},{\bf 1})_0$ &  &   \\
\hline 
\end{tabular} 
\end{center}
\caption{\sl 
The $OSp(2|4)$ representations(energy, spin, hypercharge) 
and $SU(2) \times SU(2) \times U(1)_R$ representations
in the supergravity mass spectrum for
long massive vector multiplet at
the ${\cal N}=2$ critical point and the corresponding ${\cal N}=2$
superfield in the boundary gauge theory.}
%\label{tableso4}
\end{table} 
%%%%%%%%%%%%%%%%%%%%%%%%%%%%%%%%%%%%%%%%%%%%%%%%%%%%%%%%%%%%%%%%%%%%%%%%%%%%
%%%%%%%%%%%%%%%%%%%%%%%%%%%%%%%%%%%%%%%%%%%%%%%%%%%%%%%%%%%%%%%%%%%%%%%%%%%%

%%%%%%%%%%%%%%%%%%%%%%%%%%%%%%%%%%%%%%%%%%%
%%%%%%%%%%%%%%%%%%%%%%%%%%%%%%%%%%%%%%%%%%%
$\bullet$ Short massive gravitino multiplet
%%%%%%%%%%%%%%%%%%%%%%%%%%%%%%%%%%%%%%%%%%%
%%%%%%%%%%%%%%%%%%%%%%%%%%%%%%%%%%%%%%%%%%%

The conformal dimension $\Delta$ is the twice of $U(1)_R$ charge plus
$\frac{3}{2}$ for the lowest component, $\Delta = E_0=2|R| +\frac{3}{2}$.
This  corresponds to 
spinorial superfield $\Phi_{\alpha}$ that 
satisfies $D^{+ \alpha} \Phi_{\alpha} = 0$ \cite{FFGT}. 
Of course, this constraint makes the multiplet short. 
In the $\theta^{\pm}$ expansion, the component
fields in the bulk are located with appropriate quantum numbers. 
The massless chiral superfields $\Phi_1, \Phi_2, \Phi_3, \Phi_4$
have $\Delta =\frac{1}{2}$ and $U(1)_R$ charge $\frac{1}{2}$ as
before(around (\ref{weff})). 
The gauge
superfield $W_{\alpha}$ has $\Delta =2y+ 1$ 
and $U(1)_R$ charge $y-\frac{1}{2}$ and its conjugate field has
opposite $U(1)_R$ charge $-y+\frac{1}{2}$.  
Note that the $U(1)_R$ charge for the gravitino in the $SU(3) \times
U(1)_R$
was given by $y=\frac{1}{6}$ \cite{NW}.
Then one can identify $\Tr W_{\alpha} \Phi_j$ with $({\bf 2},{\bf 2})$
and $\Tr \overline{W}_{\alpha} \overline{\Phi}_j$ with $({\bf 2},{\bf 2})$.
The corresponding $OSp(2|4)$ representations and corresponding
superfield are listed in Table 2.
For spin $\frac{3}{2}$, the relation for the mass and dimension is
given by
$\Delta =\frac{6 + |m+4|}{4}$ and for spin $0, 1$ and $\frac{1}{2}$,
the previous relations hold.

%%%%%%%%%%%%%%%%%%%%%%%%%%%%%%%%%%%%%%%%%%%%%%%%%%%%%%%%%%%%%%
%table 2%%%%%%%%%%%%%%%%%%%%%%%%%%%%%%%%%%%%%%%%%%%%%%%%%%%%%%%%%%%%
\begin{table} 
\begin{center}
\begin{tabular}{|c|c|c|c|c|c|} \hline
B.O. & Energy & Spin $0$  & Spin $\frac{1}{2}$ & Spin $1$
 & Spin $\frac{3}{2}$ \\ \hline 
$\Tr W_{\alpha} \Phi_j$ 
& $E_0
%=\frac{11}{6}
$ &  & $({\bf 2},{\bf 2})_{y}$ &  & \\
complex & $E_0 +\frac{1}{2} 
%= \frac{7}{3}
$ & 
$({\bf 2},{\bf 2})_{y- \frac{1}{2}}$  & & $({\bf 2},{\bf 2})_{y 
\pm \frac{1}{2}} $ &  \\
& $E_0+1 
%= \frac{17}{6}
$ & & $({\bf 2},{\bf 2})_{y, y- 1} $ 
&   & $({\bf 2},{\bf 2})_{y}$  \\
& $E_0+\frac{3}{2} 
%= \frac{10}{3}
$ &   &  & $({\bf 2},{\bf
  2})_{y- \frac{1}{2}}$ &  
\\
\hline
\end{tabular} 
\end{center}
\caption{\sl 
The $OSp(2|4)$ representations(energy, spin, hypercharge) 
and  $SU(2) \times SU(2)\times U(1)_R$ representations in the 
supergravity mass spectrum for
short massive gravitino multiplet at
the ${\cal N}=2$ critical point and the corresponding ${\cal N}=2$
superfield in the boundary gauge theory where $E_0= 2|y|+\frac{3}{2} =
2|R|+\frac{3}{2}$.}
%\label{tableso4}
\end{table} 
%%%%%%%%%%%%%%%%%%%%%%%%%%%%%%%%%%%%%%%%%%%%%%%%%%%%%%%%%%%%%%%%%%%%%%%%%%
%%%%%%%%%%%%%%%%%%%%%%%%%%%%%%%%%%%%%%%%%%%%%%%%%%%%%%%%%%%%%%%%%%%%%%%%%%

%%%%%%%%%%%%%%%%%%%%%%%%%%%%%%%%%%%%%%%%%%%%%%%%%%
%%%%%%%%%%%%%%%%%%%%%%%%%%%%%%%%%%%%%%%%%%%%%%%%%%
$\bullet$ ${\cal N}=2$ massless graviton multiplet
%%%%%%%%%%%%%%%%%%%%%%%%%%%%%%%%%%%%%%%%%%%%%%%%%%
%%%%%%%%%%%%%%%%%%%%%%%%%%%%%%%%%%%%%%%%%%%%%%%%%%%

This can be identified with the  
stress energy tensor superfield $T_{\alpha \beta}$
that satisfies the equations $D_{\alpha}^{\pm} T_{\alpha \beta}=0$ 
\cite{FFGRTZZ,Ahn02-3}. In components, 
the $\theta^{\pm}$ expansion of this superfield has
the stress energy tensor, the ${\cal N}=2$ supercurrents, and $U(1)_R$
symmetry current, as usual.
The conformal dimension $\Delta=2$ and the $U(1)_R$ charge is $0$.
The corresponding $OSp(2|4)$ representations and corresponding
superfield are listed in Table 3.
For spin $2$, we have the relation $\Delta_{\pm} = 
\frac{3 \pm \sqrt{9+\frac{m^2}{4}}}{2}$ and for massless case, this
leads to $\Delta_{+}=3$. This massless state can be seen from the Table 4.

%%%%%%%%%%%%%%%%%%%%%%%%%%%%%%%%%%%%%%%%%%%%%%%%%%
%%%%%%%%%%%%%%%%%%%%%%%%%%%%%%%%%%%%%%%%%%%%%%%%%%
$\bullet$ ${\cal N}=2$ massless vector multiplet
%%%%%%%%%%%%%%%%%%%%%%%%%%%%%%%%%%%%%%%%%%%%%%%%%%
%%%%%%%%%%%%%%%%%%%%%%%%%%%%%%%%%%%%%%%%%%%%%%%%%%

This conserved vector current is given by 
a scalar superfield $J^A$  
satisfying $D^{\pm \alpha} D^{\pm}_{\alpha} J^A = 0$ \cite{FFGRTZZ}. 
This transforms in the
adjoint representation of $SU(2) \times SU(2)$ 
flavor group. The corresponding boundary object
is given by 
$\Tr  \overline{\Phi}_i T^A \Phi_i$ where 
the flavor indices in $\Phi_i$ and $\overline{\Phi}_i$ are contracted
and the generator $T^A$ is   $N \times
N$ matrix  with $A=1, 2, \cdots, N^2$. 
The conformal dimension $\Delta=1$ and the $U(1)_R$ charge is $0$.
By taking a tensor product between $({\bf 2},{\bf 2})$ 
and $({\bf 2},{\bf 2})$, one
gets $({\bf 1}, {\bf 3}) \oplus ({\bf 3}, {\bf 1})$ 
of $SU(2) \times SU(2)$ representation.
The corresponding $OSp(2|4)$ representations and corresponding
superfield are listed in Table 3.

%%%%%%%%%%%%%%%%%%%%%%%%%%%%%%%%%%%%%%%%%%%%%%%%%%%%%%%%%%%%%%%%
% table 3%%%%%%%%%%%%%%%%%%%%%%%%%%%%%%%%%%%%%%%%%%%%%%%%%%%%%%%%%%%%%
\begin{table} 
\begin{center}
\begin{tabular}{|c|c|c|c|c|c|c|} \hline
B.O. & Energy & Spin $0$  & Spin $\frac{1}{2}$ & Spin $1$
 & Spin $\frac{3}{2}$ & Spin $2$ \\ \hline 
$\Tr  \overline{\Phi}_i T^A \Phi_i $ 
& $E_0
=1$ & $({\bf 1},{\bf 3} )_0 \oplus ({\bf 3},{\bf 1} )_0 $  
&  &  &  & \\
& $E_0 +\frac{1}{2} 
%= \frac{3}{2}
$ &   & $({\bf 1},{\bf 3})_{\pm\frac{1}{2}} \oplus
 ({\bf 3},{\bf 1})_{\pm\frac{1}{2}}$ &  & & \\
& $E_0+1 
%= 2
$ &  $({\bf 1},{\bf 3} )_0 \oplus ({\bf 3},{\bf 1} )_0$ &  
& $({\bf 1},{\bf 3} )_0 \oplus ({\bf 3},{\bf 1} )_0$  &  & \\
\hline
\hline
$T_{\alpha \beta}$ & $E_0=2$ & & & $({\bf 1},{\bf 1})_0$ & & \\
 & $E_0+\frac{1}{2}
%=\frac{5}{2}
$ 
& & & & $({\bf 1},{\bf 1})_{\pm\frac{1}{2}}$ & \\
& $E_0 +1 
%= 3
$ & & & & & $({\bf 1},{\bf 1})_0$ \\
\hline
\end{tabular} 
\end{center}
\caption{\sl 
The $OSp(2|4)$ representations(energy, spin, hypercharge) 
and  $SU(2) \times SU(2) \times U(1)_R$ 
representations in the supergravity mass spectrum for
``ultra'' short multiplets at
the ${\cal N}=2$ critical point and the corresponding ${\cal N}=2$
superfields in the boundary gauge theory.}
%\label{tableso4}
\end{table} 
%%%%%%%%%%%%%%%%%%%%%%%%%%%%%%%%%%%%%%%%%%%%%%%%%%%%%%%%%%%%%%%%%%%%
%%%%%%%%%%%%%%%%%%%%%%%%%%%%%%%%%%%%%%%%%%%%%%%%%%%%%%%%%%%%%%%%%%%%

We have presented the gauge invariant 
combinations of the massless superfields 
of the gauge theory whose scaling dimensions and $SU(2) \times SU(2) \times
U(1)_R$ quantum numbers exactly match the two short multiplets in
Tables $2, 3$
observed in the supergravity. 

%%%%%%%%%%%%%%%%%%%%%%%%%%%%%%%%%%%%%%%%%%%%%%%%%%%%%%%%%%%%%%%%
%%%%%%%%%%%%%%%%%%%%%%%%%%%%%%%%%%%%%%%%%%%%%%%%%%%%%%%%%%%%%%%%
\section{KK spectrum of minimally coupled scalar }
%%%%%%%%%%%%%%%%%%%%%%%%%%%%%%%%%%%%%%%%%%%%%%%%%%%%%%%%%%%%%%%%
%%%%%%%%%%%%%%%%%%%%%%%%%%%%%%%%%%%%%%%%%%%%%%%%%%%%%%%%%%%%%%%%

Let us describe the KK modes by solving the Laplace equation in
7-dimensional internal space. 
A minimally coupled scalar field 
is interacting with the gravitational field. The action 
in the background which is a warped
product of $AdS_4$ and a squashed and stretched 
7-dimensional manifold is given by 
\bea
S = \int d^{11} x \sqrt{- g} \left[ -\frac{1}{2} (\pa \phi)^2 \right].
\label{act}
\eea
The equation of motion from this action (\ref{act}) is 
\bea
\Box \phi =0.
\label{eqm}
\eea
Here $\Box$ is the 11-dimensional Laplacian. 
By exploiting the separation of variables
\bea
\phi = \hat{\Phi}(x^{\mu}, r) Y(y^m),
\label{phi}
\eea
and substituting (\ref{phi}) into (\ref{eqm}), 
one can write down (\ref{eqm}) as
\bea
Y(y^{m}) \,\Box_4 \,\hat{\Phi}(x^{\mu}, r) + \hat{\Phi}(x^{\mu}, r)\, {\cal L}
\,Y(y^{m}) =0,
\label{sol1}
\eea
where $\Box_4$ stands for the $AdS_4$ Laplacian and 
${\cal L}$ stands for a differential operator acting on  
7-dimensional manifold and is
\bea
{\cal L}  \equiv \frac{\Delta^{-1}}{\sqrt{-g_{11}}} \, \pa_{M} \left( 
\sqrt{-g_{11}} \, g_{11}^{M N} \, \pa_{N} \right)
= \frac{\Delta^{-\frac{3}{4}}}{\sqrt{g_{7}}} \, \pa_{m} \left(
   3^{-\frac{3}{2}} \, \hat{L}^{-2} \,
\Delta^{-\frac{3}{4}} \, \sqrt{g_{7}} \, g_{7}^{m n} \, \pa_{n} \right),
\label{diffop}
\eea
where $g_{mn}^7$ and $g_{MN}^{11}$ are described by the metrics (\ref{7dmet2})
and 
(\ref{11dmetric}) respectively.
The 7-dimensional metric $g_{mn}^7$ is given by explicitly
\bea
%g_{mn}^7 =
\left(
\begin{array}{ccccccc}
\frac{(2-c_{2\mu})}{3^{\frac{3}{4}}} & 0 &0 &0 &0 &0 &  0  \\ 
0 & \frac{c^2_{\mu}}{2\cdot 3^{\frac{3}{4}}} &0 &0 &0 &0 &  0 \\
0 & 0 &\frac{2c^2_{\mu}(-7+3c_{2\mu})+c_{2\theta_1} s^2_{2\mu}}
{16\cdot 3^{\frac{3}{4}} (-2 +c_{2\mu}) } &0
&\frac{c_{\theta_1}c_{\theta_2}
c^2_{\mu}(-3+c_{2\mu})}{4\cdot3^{\frac{3}{4}}(-2+c_{2\mu})} & 
\frac{c_{\theta_1}
c^2_{\mu}(-3+c_{2\mu})}{4\cdot3^{\frac{3}{4}}(-2+c_{2\mu})}  &
\frac{c_{\theta_1} c^2_{\mu}}{2\cdot 3^{\frac{3}{4}}} \\
0 & 0 &0 & \frac{c^2_{\mu}}{2\cdot 3^{\frac{3}{4}}}
& 0 & 0 &  0 \\
0 & 0 & \frac{c_{\theta_1} c_{\theta_2}
c^2_{\mu}(-3+c_{2\mu})}{4\cdot3^{\frac{3}{4}}(-2+c_{2\mu})}  & 0
 & \frac{2 
c^2_{\mu}(-7+3 c_{2\mu}) + c_{2\theta_2}
s^2_{2\mu}}{16\cdot3^{\frac{3}{4}}
(-2+c_{2\mu})} 
&  \frac{c_{\theta_2}
c^2_{\mu}(-3+c_{2\mu})}{4\cdot3^{\frac{3}{4}}(-2+c_{2\mu})} &
\frac{c_{\theta_2}
c^2_{\mu}}{2\cdot3^{\frac{3}{4}}}   
\\
0 & 0 &\frac{c_{\theta_1}
c^2_{\mu}(-3+c_{2\mu})}{4\cdot3^{\frac{3}{4}}(-2+c_{2\mu})}  & 
0
& \frac{c_{\theta_2}
c^2_{\mu}(-3+c_{2\mu})}{4\cdot3^{\frac{3}{4}}(-2+c_{2\mu})}  &
\frac{
c^2_{\mu}(-3+c_{2\mu})}{4\cdot3^{\frac{3}{4}}(-2+c_{2\mu})}  &  \frac{
c^2_{\mu}}{2\cdot3^{\frac{3}{4}}}  \\
0 & 0 & \frac{c_{\theta_1}
c^2_{\mu}}{2\cdot3^{\frac{3}{4}}}  &0 &\frac{c_{\theta_2}
c^2_{\mu}}{2\cdot3^{\frac{3}{4}}}  & \frac{
c^2_{\mu}}{2\cdot3^{\frac{3}{4}}}  &   
\frac{(2-c_{2\mu})}{2\cdot3^{\frac{3}{4}}} 
\end{array} \right),
\label{7dmetric}
\eea
where we use the simplified notation $c_{2\mu}\equiv \cos 2\mu$
and so on and let us introduce the
angular coordinates $y^m \equiv (\mu, \theta_1, \phi_1, \theta_2,
\phi_2, \psi, \phi)$.

Let us see what is the eigenfunction $Y(y^m)$ of the differential operator 
${\cal L}$
\bea
{\cal L} \, Y(y^m) = -m^2 \, Y(y^m).
\label{diff1}
\eea
Then the equation (\ref{sol1}) implies the equation of motion of a
massive scalar field in $AdS_4$:
\bea
 \Box_4 \, \hat{\Phi}(x^{\mu}, r) -m^2 \, \hat{\Phi}(x^{\mu}, r) = 0. 
\label{ads4}
\eea
Therefore, one obtains a
tower of KK modes which are all massive scalars (\ref{ads4}) with
masses 
$m^2$ determined
by the eigenvalues of the above differential operator ${\cal L}$. 

The spin-2 massive ${\cal N}=8$ supermultiplet \cite{DNP} 
at level $n$ is described
by the $SO(8)$ Dynkin labels $(n,0,0,0)$, this breaks into the $SO(7)$
Dynkin labels $(0,0,n)$, and finally the massive multiplets of ${\cal
  N}=8$ for $n=1,2, \cdots$, are decomposed into the various representations
under the $SU(2) \times SU(2)$ symmetry. In particular, one has, with
the help of \cite{ps,OT},  
\bea
SO(8) & \rightarrow & SO(7)   \rightarrow SU(4) \rightarrow SU(2)^2, \nonu \\
{\bf 8}_v(1,0,0,0) & \rightarrow & {\bf 8}(0,0,1)  \rightarrow {\bf 4}
\oplus \overline{\bf 4}  \rightarrow 2 ({\bf
  2}, {\bf 1}) \oplus
2 ({\bf 1},{\bf 2}),
\nonu \\
{\bf 35}_v(2,0,0,0) &\rightarrow & {\bf 35}(0,0,2) \rightarrow  {\bf
  10} \oplus \overline{\bf 10} \oplus {\bf 15}   \rightarrow 
({\bf 1},{\bf 1})   
\oplus  4 ({\bf 2},{\bf 2}) \oplus 3
({\bf 3},{\bf
1})  \oplus  3 ({\bf 1},{\bf 3}), \nonu \\
{\bf 112}_v(3,0,0,0) & \rightarrow & {\bf 112}'(0,0,3) \rightarrow
{\bf 20''} \oplus \overline{\bf 20}'' \oplus {\bf 36} \oplus
\overline{\bf 36} \nonu \\
& \rightarrow & 
2 ({\bf 2},{\bf 1})   
\oplus  2 ({\bf 1},{\bf 2}) \oplus  4 ({\bf 4},{\bf 1}) \oplus 6
({\bf 3},{\bf 2}) \oplus  6 ({\bf 2},{\bf 3}) \oplus  4 ({\bf 1},{\bf 4}).
\label{sobranching} 
\eea

The differential operator acting on 7-dimensional manifold 
is given by (\ref{diffop}) and this can be rewritten, from the metric
(\ref{7dmetric}) and the warp factor (\ref{delta}) with (\ref{rhochi}) and
(\ref{X}),  in terms of angular
coordinates
as follows:
\bea
{\cal L}  & = &  -\frac{1}{6\, \hat{L}^2} 
(-2 +\cos2\mu) \sec^2\mu \,{\cal C}_2
 + \frac{1}{2 \, \hat{L}^2}\left[
\pa_{\mu}^2 +(\cot\mu-5 \tan\mu) \, \pa_{\mu} \right. \nonu \\
 & + & \left.  (1+\csc^2\mu) \,
\pa_{\psi}^2 +(2+\csc^2\mu)\, \pa_{\phi}^2 -2(2+\csc^2\mu)\, \pa_{\psi}
\pa_{\phi} \right],
\label{calL}
\eea
where the quadratic differential operator can be written as 
\bea
{\cal C}_2 \equiv
6 \sum_{i=1}^2
\left[\frac{1}{\sin \theta_i} \, \pa_{\theta_i} \, \sin \theta_i \, \pa_{\theta_i} +
\left( \frac{1}{\sin \theta_i} \, 
\pa_{\phi_i} -\cot \theta_i \, \pa_{\phi} \right)^2\right]
+ 9 \pa_{\phi}^2. 
\label{c2diff}
\eea
What are the corresponding eigenfunctions?
It is known that the scalar spherical harmonic for $D$-sphere 
${\bf S}^D$ is described
by each independent component of totally symmetric traceless tensor of
rank $n$. 
Also one can write down the eigenfunctions via the
separation of variables as follows: 
\bea
Y(y^m) = J_1(\theta_1) \,e^{i m_1 \phi_1} \,J_2(\theta_2) \,e^{i m_2 \phi_2}
\,e^{\frac{i}{2} R_{\phi} \phi} \,e^{\frac{i}{2} R_{\psi} \psi} \, H(\cos^2 \mu). 
\label{y}
\eea
Then the solution for $ J_1(\theta_1)$ is a linear combination of 
the following two independent hypergeometric functions \cite{KM}
\bea
j_1(\theta_1) & = & \sin^{m_1} \theta_1 \cot^{\frac{R_{\phi}}{2}} 
(\frac{\theta_1}{2})
\,\HGF{-l_1+m_1}{1+l_1+m_1}{1+m_1-
\frac{R_{\phi}}{2}}{\sin^2 \frac{\theta_1}{2}},
\nonu \\
j_2(\theta_1) & = & \sin^{\frac{R_{\phi}}{2}} \theta_1 \cot^{m_1} 
(\frac{\theta_1}{2})
\, \HGF{-l_1+\frac{R_{\phi}}{2}}{1+l_1+\frac{R_{\phi}}{2}}{1-m_1+
\frac{R_{\phi}}{2}}{\sin^2 \frac{\theta_1}{2}}.
\label{JJs}
\eea
When $m_1 \leq \frac{R_{\phi}}{2}$, the function $j_2(\theta_1)$ is
non-singular while when  $m_1 \geq \frac{R_{\phi}}{2}$, 
the function $j_1(\theta_1)$ is non-singular.
Similarly, the solution for $ J_2(\theta_2)$ is also a linear combination of 
these hypergeometric functions by replacing $\theta_1, m_1, l_1, R_{\phi}$ with
$\theta_2, m_2, l_2, R_{\psi}$ respectively.
The relevant eigenvalues for the above quadratic differential operator (\ref{c2diff})
can be computed explicitly as follows \cite{Gubser,CDDF}:
\bea
&& {\cal C}_2 \, J_1(\theta_1) \,e^{i m_1 \phi_1} \,J_2(\theta_2) \,e^{i m_2 \phi_2}
\,e^{\frac{i}{2} R_{\phi} \phi} \nonu \\
&& = \left[-6l_1(l_1+1) -
6l_2(l_2+1) +\frac{3}{4} R_{\phi}^2 \right]
J_1(\theta_1) \,e^{i m_1 \phi_1} \,J_2(\theta_2) \,e^{i m_2 \phi_2}
\,e^{\frac{i}{2} R_{\phi} \phi}.
\label{c2eigenvalue}
\eea

It turns out that 
the eigenvalue problem (\ref{diff1}) 
leads to the following nontrivial differential equation
\bea
(1-u) \, u \, H''(u) + (3- 4u) \, H'(u) + \left(A +\frac{B}{u-1}
  +\frac{C}{u} \right) 
H(u) =0, 
\qquad u \equiv \cos^2 \mu,
\label{diff}
\eea
where we introduce the following quantities 
\bea
A & \equiv & l_1(l_1+1)+l_2(l_2+1)-\frac{1}{4} 
\left( R_{\phi}-\frac{R_{\psi}}{2} \right)^2 + \frac{1}{2} m^2 \hat{L}^2,
\nonu \\
B &\equiv & \frac{1}{16} (R_{\phi}-R_{\psi})^2, \qquad
C \equiv - 
\frac{3}{2}\left[l_1(l_1+l_1)+l_2(l_2+1) \right] + \frac{3}{16} R_{\phi}^2.
\label{ABC}
\eea
Note that the contribution from  the ${\cal C}_2$ term in 
(\ref{calL}) occurs only in the
term of $H(u)$ in (\ref{diff}) while the contributions from other terms 
in (\ref{calL})  occur all the terms in $H''(u), H'(u)$ or $H(u)$ of (\ref{diff}). 
In general, the solutions for (\ref{diff}) can be written in terms of
two independent hypergeometric functions but they are rather
complicated due to the fact that the linear terms in $H(u)$ of
(\ref{diff}) depend on the variable $u$, compared with \cite{KPR,AW0907}. 
By introducing the R-charge which is a linear combination between 
$R_{\phi}$ and $R_{\psi}$ with appropriate normalization from (\ref{R}), 
\bea
 R \equiv  \frac{1}{2}\left(R_{\phi}-\frac{R_{\psi}}{2}\right),
\label{defR}
\eea
and requiring that $R_{\phi}=R_{\psi}$ which leads to the vanishing of
$B$ in (\ref{ABC}),
one obtains the solutions in very simple form.
The relative sign difference in (\ref{defR}) is due to the fact that 
the combination $\phi + 2\psi$ should be invariant quantity under the
two
$U(1)$ symmetries, stressed in section 2.
Then the KK spectrum 
of minimally coupled scalar can be obtained by putting the first
argument of the hypergeometric function to be negative integer or zero $-j$
and solving for $m^2$. Then the mass-squared in $AdS_4$ can be
written, in terms of $l_1, l_2, R$ and $j$, as
\bea
 m^2  
=  
 \frac{1}{\hat{L}^2}\left[2j^2 +2 j(1+  \sqrt{
4 - {\cal C}_2})-\frac{1}{6} {\cal C}_2 + \sqrt{4- {\cal C}_2}  -2 R^2  
-2  \right].
\label{square}
\eea
Here ${\cal C}_2$ is an eigenvalue appearing in (\ref{c2eigenvalue})
$ {\cal C}_2 \equiv -6l_1(l_1+1) -
6l_2(l_2+1) +12 R^2 $.
Then the regular solution for (\ref{diff}) is given by
\bea
H(u) =
u^{\sqrt{4-{\cal C}_2}-1} \, \HGF{-j}{j+1+
\sqrt{4-{\cal C}_2}}{1+\sqrt{4-{\cal C}_2}}{u}, \qquad u \equiv \cos^2
\mu.
\label{hu}
\eea
Substituting this into the (\ref{y}), one gets 
the eigenfunctions of (\ref{diff1}).
For $j=0$, the hypergeometric function becomes $u$-independent
constant and the contribution from $H(u)$ arises in $u^{\sqrt{4-{\cal C}_2}-1}$.
For $j=1$, the hypergeometric function leads to
$1-\frac{2+\sqrt{4-{\cal C}_2}}{1+\sqrt{4-{\cal C}_2}} u$.
For general nonzero $j$, the hypergeometric function is a polynomial
of order $j$ in $u$.

The dimension of the CFT operators dual to the KK modes 
can be obtained from the AdS/CFT correspondence \cite{Maldacena}
\bea
\Delta(\Delta-3) = m^2 \hat{L}^2.
\label{Del}
\eea
The $OSp(2|4)$ supermultiplets with spin-2 components  
are massless graviton multiplet with
$SD(2,1,0|2)$ denoted by \cite{Merlatti}, 
short  graviton multiplet with $SD(y_0+2, 1, y_0|2)$ where $y_0 > 0$
and long  graviton multiplet with
$SD(E_0, 1, y_0|2)$ where $E_0 > y_0+ 2$ and $y_0 \geq 0$.
As recognized in \cite{Ahn0806n2}, the massless graviton multiplet has
conformal dimension $\Delta=3$(the ground state component has
dimension $\Delta_0=2$ and see the Table 3 of previous section or 
the Table 5 of \cite{Ahn0806n2}).
This ${\cal N}=2$ massless graviton multiplet decomposes into 
$SD(\frac{5}{2},\frac{3}{2}|1) \oplus SD(2,1|1)$ of ${\cal N}=1$. 
Similarly ${\cal N}=2$ short graviton multiplet $SD(y_0
+2, 1, y_0|2)$ decomposes into 
$ SD(y_0+\frac{5}{2},\frac{3}{2}|1) \oplus
 SD(y_0+2,1|1)$. Since $y_0>0$, the spin-2 component of this
multiplet has 
conformal dimension $\Delta = y_0+3 > 3$ and this gives massive modes
according to (\ref{Del}).
The ${\cal N}=2$ long graviton multiplet $SD(E_0, 1, y_0|2)$ decomposes into 
$ SD(E_0+\frac{1}{2},\frac{3}{2}|1) \oplus
 SD(E_0+1,1|1) \oplus  SD(E_0+\frac{1}{2},\frac{1}{2}|1)\oplus  SD(E_0,1|1)$.
Also  the spin-2 component of this
multiplet has 
conformal dimension $\Delta = E_0+1 > y_0+3 > 3$ and this provides
massive 
modes due to (\ref{Del}).

The gauge theory conjectured to be dual to the $SU(2) \times SU(2)
\times U(1)_R$ ${\cal N }=2$ supergravity background is a 
deformation of  the quiver diagram for a partial resolution of
$Q^{1,1,1}$ theory
by a superpotential  term quadratic in
$\Phi$ (\ref{quad}) which is an adjoint field. The gauge theory has also $SU(2)
\times SU(2) \times U(1)_R$ symmetry where the $SU(2) \times SU(2)$
symmetry corresponds to the global rotations of ${B}_1, {B}_2$
and ${C}_1, {C}_2$ in (\ref{b1b2c1c2}). Under the $U(1)_R$
symmetry, the four fields as well as an adjoint field 
have R-charges given by 
\bea
&& R({B}_1)  =  R({B}_2)=R({C}_1)=R({C}_2)
=\frac{1}{4},
\qquad 
R(\Phi) = 1, \nonu \\
&& R(\overline{B}_1)  =  R(\overline{B}_2)=R(\overline{C}_1)=R(\overline{C}_2)
=-\frac{1}{4},
\qquad 
R(\overline{\Phi}) = -1,
\label{fieldRcharge}
\eea
as in previous section, around (\ref{weff}).

In Tables 4 and 5, we present a few of these
modes and also provide the dual gauge theory
operators. 
The branching rule for $SO(8)$ into $SU(2) \times SU(2) \times U(1)_R$ 
is given by (\ref{sobranching}).
The quantum numbers $l_1$ and $l_2$ of $SU(2) \times SU(2) \times
U(1)_R$ 
is characterized by each representation  $({\bf 2l_1+1})$ and $({\bf
  2l_2+1})$ 
as before. The KK excitation mode 
$j$ is nonnegative integer for the finiteness of hypergeometric
function. 
The conformal dimension of dual SCFT operator is given by 
(\ref{Del}). As the mass-squared formula is used from
(\ref{square}), then this conformal dimension is determined.
Starting with the ${\cal N}=2$ SCFT operator denoted by
$T_{\alpha \beta}$ 
corresponding to the massless graviton multiplet we introduced in
previous section, one constructs a
tower of KK modes by multiplying ${B}_i$ or ${C}_j$(and its
conjugated fields) with $T_{\alpha \beta}$ in addition to the overall
factor which depends on $\Phi$(and $\overline{\Phi}$), 
for $j=0$ modes. For nonzero $j$'s, 
there appear some polynomials in $\Phi \overline{\Phi}$.  

It is natural to identify, up to normalization, these fields ${B}_i$
and 
${C}_j$ and the adjoint field $\Phi$, that are noncommuting operators
in the gauge theory side, with 
the angular coordinates as follows: 
\bea
{B}_1 & \leftrightarrow & \cos (\frac{\theta_1}{2}) \, 
e^{\frac{i}{2}(\phi+\phi_1)}, 
\qquad 
{B}_2  \leftrightarrow \sin (\frac{\theta_1}{2}) \, 
e^{\frac{i}{2}(\phi-\phi_1)},
\nonu \\
{C}_1 &  \leftrightarrow &  \cos (\frac{\theta_2}{2}) \, 
e^{\frac{i}{2}(\psi+\phi_2)}, 
\qquad 
{C}_2 \leftrightarrow \sin (\frac{\theta_2}{2}) \, 
e^{\frac{i}{2}(\psi-\phi_2)},
\nonu \\
\Phi & \leftrightarrow & \sin \mu \, e^{-i\psi}.
\label{fivefields}
\eea
Actually first four of them can be read off from  (\ref{zis})
in the context of conifold description. The last one in
(\ref{fivefields})
can be read off from (\ref{w}). If we shift the fields by
(\ref{shift}), then the corresponding objects also arise in the Tables 4 and 5.
Note that the factor $\sqrt{\cos \mu}$ corresponds to 
$(1-\Phi \overline{\Phi})^{\frac{1}{4}}$ in the ${\cal N}=2$ boundary field theory.

One can see these features from the solutions of Laplacian eigenvalue
equation we have found. From the explicit form in (\ref{JJs}),
one computes the following expressions, by choosing the appropriate regular
solutions, 
for given quantum numbers 
\bea
j_1(\theta_1)|_{m_1=\frac{1}{2},l_1=\frac{1}{2},R_{\phi}=1} & \sim & \cos
(\frac{\theta_1}{2}),
\qquad
j_2(\theta_1)|_{m_1=-\frac{1}{2},l_1=\frac{1}{2},R_{\phi}=1} \sim \sin
(\frac{\theta_1}{2}),
\nonu \\
j_1(\theta_2)|_{m_2=\frac{1}{2},l_2=\frac{1}{2},R_{\psi}=1} & \sim & \cos
(\frac{\theta_2}{2}),
\qquad
j_2(\theta_2)|_{m_2=-\frac{1}{2},l_2=\frac{1}{2},R_{\psi}=1} \sim \sin
(\frac{\theta_2}{2}).
\label{jjjj}
\eea
Then it is easy to see that these four solutions of (\ref{jjjj}) appear 
in the right hand sides of (\ref{fivefields}) exactly.
Therefore, the eigenfunctions 
except the function $H(\cos^2 \mu)$ in (\ref{y}) correspond to 
${B}_i$ or ${C}_j$ and its conjugated fields.
Note that the exponential factors of (\ref{y}) 
already appear in the chiral fields given by 
(\ref{fivefields}).

Let us describe the ${\cal N}=2$ SCFT operators in Tables 4 and 5.
In the representations $({\bf 2}, {\bf 1})_{\pm \frac{1}{4}}$ and 
$({\bf 1}, {\bf 2})_{\pm \frac{1}{4}}$ coming from ${\bf 8}_v$ of
$SO(8)$
representation, one can read off  the corresponding 
${\cal N}=2$ SCFT operators by using the fact that each 
${B}_i$ and ${C}_j$ consists of a doublet of each $SU(2)$ 
group and the $\mu$-dependent part can be determined by (\ref{hu}) for
given quantum numbers. For example, $l_1=\frac{1}{2}, m_1=\pm \frac{1}{2},
R_{\phi}=\pm 1=4R$ corresponding to  $({\bf 2}, {\bf 1})_{\pm \frac{1}{4}}$. 
Recall that $u \equiv \cos^2 \mu$ corresponds
to $(1-\Phi\overline{\Phi})$ and the eigenvalue ${\cal C}_2$ does not
depend on the quantum number $m_1$ or $m_2$.
Since the KK excitation number vanishes $j=0$, the $\mu$ dependence arises from
only $u^{\sqrt{4-{\cal C}_2}-1}$.  
Note that the eigenvalue (\ref{c2eigenvalue}) has a
symmetry under the $l_1 \leftrightarrow l_2$ or under the $R 
\leftrightarrow -R$. 

At the next level, the additional structure in ${\cal N}=2$ SCFT
operator, compared to the $j=0$ case,  
corresponding to the representation $({\bf 1},{\bf 1})_0$ with
nonzero $j=1$ can be constructed from (\ref{hu}) by substituting $j=1$.
In this case, we should go to the differential equation (\ref{diff})
directly in order to obtain the correct solution.
The solution is given by $H(u) = \HGF{\frac{1}{2}[3-\sqrt{9+
2 m^2 \hat{L}^2}]}{\frac{1}{2}[3+\sqrt{9+ 2m^2 \hat{L}^2}]}{3}{u}$.
Now we move on the other representations $({\bf 3}, {\bf 1})_{\pm
  \frac{1}{2},0}$.
Since $({\bf 3},{\bf 1})$ transforms as an adjoint representation of
the first $SU(2)$,
this can be constructed by a tensor product between $({\bf 2},{\bf
  1})$ and $({\bf 2},{\bf 1})$ which becomes $({\bf 3},{\bf 1}) \oplus
({\bf 1},{\bf 1})$. 
Depending on the correct R-charges 
(\ref{fieldRcharge}),
one determines the appropriate combinations of ${B}_i$ and 
$\overline{B}_j$.  Moreover, the $\mu$-dependent factor 
can be determined by (\ref{hu}) for
given quantum numbers. For example, let us consider the quantum
numbers $l_1=1, l_2=0,
R_{\phi}=\pm 2, 0=4R$.  For the representations 
$({\bf 2},{\bf
  2})_{\pm \frac{1}{2}, 0,0}$, one sees that 
the tensor product between $({\bf 2},{\bf
  1})$ and $({\bf 1},{\bf 2})$ provides these particular cases.
In this case,  the exponent of $\mu$-dependent factor 
appears in two different values by (\ref{hu}) depending on vanishing
R-charge (\ref{defR}) or non-vanishing R-charge for
given quantum numbers, i.e., $l_1=\frac{1}{2}=l_2,
R_{\phi}=\pm 2, 0=4R$. Finally, the analysis for the representations 
$({\bf 1}, {\bf 3})_{\pm
  \frac{1}{2},0}$ can be done similarly, as in $({\bf 3}, {\bf 1})_{\pm
  \frac{1}{2},0}$.

%%%%%%%%%%%%%%%%%%%%%%%%%%%%%%%%%%%%%%%%%%%%%%%%%%%%%%%%%%%%%%%%%%%%%%
%table 4%%%%%%%%%%%%%%%%%%%%%%%%%%%%%%%%%%%%%%%%%%%%%%%%%%%%%%%%%%%%%%%%%%%%%
\begin{table} 
\begin{center}
\begin{tabular}{|c|c|c|c|c|c|c|} \hline
$SO(8)$  & $SU(2)^2_R$ & $R_{\phi}$& $R_{\psi}$
& $j$  
%& $\Delta$ 
& $m^2 \hat{L}^2$ & ${\cal N}=2$ \mbox{SCFT Operator}  \\
\hline
${\bf 1}$ &   
$({\bf 1},{\bf 1})_0$ & $0$ & $0$ & $0$  
%& $3$  
& $0$ & 
$T_{\alpha \beta}$ \\
\hline 
${\bf 8}_v$& $({\bf 2},{\bf 1})_{+\frac{1}{4}}$ &$1$ &$1$&  $0$ 
%&
%$\frac{1}{2}(3+
%\sqrt{3+2\sqrt{31}})$ 
&
$\frac{1}{2}(\sqrt{31}-3)$ 
& $T_{\alpha \beta} {B}_i(1-\Phi\overline{\Phi})^
{\frac{\sqrt{31}}{4}-1}$ \\
& $({\bf 2},{\bf 1})_{-\frac{1}{4}}$ & $-1$ &$-1$ & $0$ 
%& $\frac{1}{2}(3+
%\sqrt{3+2\sqrt{31}})$ 
&
$\frac{1}{2}(\sqrt{31}-3)$ & $T_{\alpha \beta}
\overline{B}_i(1-\Phi\overline{\Phi})^
{\frac{\sqrt{31}}{4}-1}
$ \\
&  
$({\bf 1},{\bf 2})_{+\frac{1}{4}}$ &$1$ & $1$ & $0$  
%& $\frac{1}{2}(3+
%\sqrt{3+2\sqrt{31}}) $ 
& $\frac{1}{2}(\sqrt{31}-3)$ & 
$T_{\alpha \beta} {C}_i(1-\Phi\overline{\Phi})^
{\frac{\sqrt{31}}{4}-1}$ \\
&  $({\bf 1},{\bf 2})_{-\frac{1}{4}}$ & $-1$ &$-1$ & $0$  
%& $\frac{1}{2}(3+
%\sqrt{3+2\sqrt{31}}) $
&$\frac{1}{2}(\sqrt{31}-3)$ 
& $T_{\alpha \beta}\overline{C}_i(1-\Phi\overline{\Phi})^
{\frac{\sqrt{31}}{4}-1} $ \\
\hline
$ {\bf 35}_v$ 
&  $({\bf 1},{\bf 1})_0$ & $0$ & $0$ & $1$ 
%& $\frac{1}{2}(3+\sqrt{41})$  
& $8$ 
& $T_{\alpha \beta}( -1+ 4 \Phi\overline{\Phi})$ \\
& $({\bf 3},{\bf 1})_{+\frac{1}{2}}$ &$2$& $2$ & $0$ 
%& $\frac{1}{2}(3+
%\sqrt{5+4\sqrt{13}})$ 
& $\sqrt{13}-1$ & $T_{\alpha \beta}({B}_i {B}_j -
\frac{1}{2} \delta_{ij} {B}_k {B}_k)(1-\Phi\overline{\Phi})^
{\frac{\sqrt{13}}{2}-1} $  \\
& $({\bf 3},{\bf 1})_0$ &$0$& $0$ & $0$ 
%& $4$ 
& $4$ & $T_{\alpha \beta}({\cal
  B}_i \overline{B}_j -\frac{1}{2} \delta_{ij} {B}_k \overline{B}_k)
(1-\Phi\overline{\Phi}) $  \\
& $({\bf 3},{\bf 1})_{-\frac{1}{2}}$ &$-2$&$-2$ & $0$ 
%& $\frac{1}{2}(3+
%\sqrt{5+4\sqrt{13}})$ 
& $\sqrt{13}-1 $ & $T_{\alpha \beta}(\overline{B}_i 
\overline{B}_j -\frac{1}{2} \delta_{ij} \overline{B}_k \overline{B}_k)
(1-\Phi\overline{\Phi})^
{\frac{\sqrt{13}}{2}-1} $  \\
&  $({\bf 2},{\bf 2})_{+\frac{1}{2}}$ &$2$& $2$ & $0$ 
%&$\frac{1}{2}(3+\sqrt{3+4\sqrt{10}}) $ 
& $\sqrt{10}-\frac{3}{2}$ &  $T_{\alpha \beta} {B}_i 
{C}_j(1-\Phi\overline{\Phi})^
{\frac{\sqrt{10}}{2}-1}$ 
%-\frac{1}{2} \delta_{ij} {B}_k {\cal
%C}_k)$
\\
&  $({\bf 2},{\bf 2})_0$ &$0$& $0$ & $0$ 
%&$\frac{1}{2}(3+\sqrt{7+4\sqrt{13}}) $ 
& $\sqrt{13}-\frac{1}{2}$ &  $T_{\alpha \beta} {B}_i 
\overline{C}_j(1-\Phi\overline{\Phi})^
{\frac{\sqrt{13}}{2}-1} $
%-\frac{1}{2} \delta_{ij} {B}_k \overline{\cal
%C}_k)$
\\
&  $({\bf 2},{\bf 2})_0$ &$0$& $0$ & $0$ 
%&$\frac{1}{2}(3+\sqrt{7+4\sqrt{13}}) $ 
& $\sqrt{13}-\frac{1}{2}$ &  $T_{\alpha \beta}
\overline{B}_i 
{C}_j(1-\Phi\overline{\Phi})^
{\frac{\sqrt{13}}{2}-1}$ 
%-\frac{1}{2} \delta_{ij} \overline{B}_k {\cal
%C}_k)$
\\
&  $({\bf 2},{\bf 2})_{-\frac{1}{2}}$ &$-2$& $-2$ & $0$ 
%&$\frac{1}{2}(3+\sqrt{3+4\sqrt{10}})$ 
& $\sqrt{10}-\frac{3}{2}$ &  $T_{\alpha \beta}
\overline{B}_i 
\overline{C}_j(1-\Phi\overline{\Phi})^
{\frac{\sqrt{10}}{2}-1} $
%-\frac{1}{2} \delta_{ij} \overline{\cal
%  B}_k 
%\overline{\cal
%C}_k)$  
\\
&  $({\bf 1},{\bf 3})_{+\frac{1}{2}}$ &$2$& $2$ & $0$ 
%&$\frac{1}{2}(3+
%\sqrt{5+4\sqrt{13}}) $ 
& $\sqrt{13}-1$ &  $T_{\alpha \beta}({C}_i 
{C}_j -\frac{1}{2} \delta_{ij} {C}_k {C}_k)(1-\Phi\overline{\Phi})^
{\frac{\sqrt{13}}{2}-1}$
\\
&  $({\bf 1},{\bf 3})_0$ &$0$ & $0$ & $0$ 
%&$4 $ 
& $4$ &  $T_{\alpha \beta}({C}_i 
\overline{C}_j -\frac{1}{2} \delta_{ij} {C}_k 
\overline{C}_k)(1-\Phi\overline{\Phi})$
\\
&  $({\bf 1},{\bf 3})_{-\frac{1}{2}}$ &$-2$ &$-2$ & $0$ 
%&$\frac{1}{2}(3+
%\sqrt{5+4\sqrt{13}}) $ 
& $\sqrt{13}-1$ &  $T_{\alpha \beta}(\overline{C}_i 
\overline{C}_j -\frac{1}{2} \delta_{ij} 
\overline{C}_k \overline{C}_k)(1-\Phi\overline{\Phi})^
{\frac{\sqrt{13}}{2}-1}$
\\
\hline
\end{tabular} 
\end{center}
\caption{\sl 
The first few spin-2 components of the massive(and massless) 
graviton multiplets. For each
multiplet we present $SO(8)$ representation (\ref{sobranching}), $SU(2)
\times SU(2) \times U(1)_{R}$
representations denoted by $({\bf 2l_1+1},{\bf 2l_2+1 })_R$,
$R_{\phi}$-charge, $R_{\psi}$-charge, the KK
excitation number $j$, 
%the dimension $\Delta$ (\ref{Del}) of 
%the spin-2 component
%of the multiplet, 
the mass-squared $m^2 \hat{L}^2$ (\ref{square}) 
of the $AdS_4$ field
and the corresponding ${\cal N}=2$ dual SCFT operator.
The dimension $\Delta$ (\ref{Del}) of the spin-2 component
of the multiplet we do not present here due to 
the space can be obtained also.}
%\label{tableso4}
\end{table} 
%%%%%%%%%%%%%%%%%%%%%%%%%%%%%%%%%%%%%%%%%%%%%%%%%%%%%%%%%%%%%%%%%%%%%%%%%%%%
%%%%%%%%%%%%%%%%%%%%%%%%%%%%%%%%%%%%%%%%%%%%%%%%%%%%%%%%%%%%%%%%%%%%%%%%%%%%

At the next level, 
the extra structure in ${\cal N}=2$ SCFT
operator, compared to $j=0$ case,  corresponding to the representations 
$({\bf 2},{\bf 1})_{\pm \frac{1}{4}}$ and 
$({\bf 1},{\bf 2})_{\pm \frac{1}{4}}$ with
nonzero $j=1$ can be analyzed from (\ref{hu}) by 
substituting $j=1$ from the hypergeometic function.
Let us move on the other representations $({\bf 4}, {\bf 1})_{\pm
  \frac{3}{4}, \pm \frac{1}{4}}$.
Since $({\bf 4},{\bf 1})$ transforms as a symmetric representation of
the first $SU(2)$,
this can be determined by a tensor product between $({\bf 3},{\bf
  1})$ and $({\bf 2},{\bf 1})$ which leads to $({\bf 4},{\bf 1}) \oplus
({\bf 2},{\bf 1})$. 
Depending on the correct R-charges,
one can determine the appropriate triple-product 
combinations of ${B}_i$ and 
$\overline{B}_j$.  Moreover, the $\mu$-dependent factor 
can be determined by (\ref{hu}) for
given quantum numbers, i.e., $l_1=\frac{3}{2}, l_2=0,
R_{\phi}=\pm 3, \pm 1=4R$.
For the representations 
$({\bf 3},{\bf
  2})_{\pm \frac{3}{4}, \pm \frac{1}{4}, \pm \frac{1}{4}}$, 
the tensor product between $({\bf 3},{\bf
  1})$ and $({\bf 1},{\bf 2})$ provides these cases: the product
between the adjoint from
the first $SU(2)$ and the fundamental from the second $SU(2)$.
The exponent of $\mu$-dependent factor 
appears in two different values by (\ref{hu}) depending on $\pm \frac{3}{4}$
R-charges or $\pm \frac{1}{4}$  R-charges for
given quantum numbers, i.e., $l_1=1, l_2=\frac{1}{2},
R_{\phi}=\pm 3, \pm 1$.
Finally, the analysis for the representations 
$({\bf 2}, {\bf 3})_{\pm
  \frac{3}{4},\pm \frac{1}{4}, \pm\frac{1}{4}}$ and  $({\bf 4}, {\bf 1})_{\pm
  \frac{3}{4}, \pm \frac{1}{4}}$ can be done similarly by using the
 symmetry between the quantum numbers.

%%%%%%%%%%%%%%%%%%%%%%%%%%%%%%%%%%%%%%%%%%%%%%%%%%%%%%%%%%%%%%%%%%%%%%
%table 5%%%%%%%%%%%%%%%%%%%%%%%%%%%%%%%%%%%%%%%%%%%%%%%%%%%%%%%%%%%%%%%%%%%%%
\begin{table} 
\begin{center}
\begin{tabular}{|c|c|c|c|c|c|c|} \hline
$SO(8)$  & $SU(2)^2_R$ & $R_{\phi}$ & $R_{\psi}$
& $j$  
%& $\Delta$ 
& $m^2 \hat{L}^2$ & ${\cal N}=2$ \mbox{SCFT Operator}  \\
\hline
${\bf 112}_v$ 
& $({\bf 2},{\bf 1})_{+\frac{1}{4}}$ &$1$ &$1$ & $1$ 
%&
%$\frac{1}{2}( 3+ \sqrt{19+6\sqrt{31}})$  
& $\frac{1}{2}(3\sqrt{31}+5)$ & ${T}_{\alpha \beta}
{B}_i(1-\Phi \overline{\Phi})^{\frac{\sqrt{31}}{4}-1}
[ 1-  (2+\frac{\sqrt{31}}{2})\Phi\overline{\Phi}]$  \\
& $({\bf 2},{\bf 1})_{-\frac{1}{4}}$ &$-1$ & $-1$ & $1$ 
%& $\frac{1}{2}( 3+ \sqrt{19+6\sqrt{31}})$  
& $\frac{1}{2}(3\sqrt{31}+5)$ & $T_{\alpha \beta}
\overline{B}_i (1-\Phi \overline{\Phi})^{\frac{\sqrt{31}}{4}-1}
[ 1-  (2+\frac{\sqrt{31}}{2})\Phi\overline{\Phi}]$  \\
& $({\bf 1},{\bf 2})_{+\frac{1}{4}}$ &$1$ & $1$ & $1$ 
%& $\frac{1}{2}( 3+ \sqrt{19+6\sqrt{31}})$ 
 & $\frac{1}{2}(3\sqrt{31}+5)$ & $T_{\alpha \beta}
{C}_i (1-\Phi \overline{\Phi})^{\frac{\sqrt{31}}{4}-1}
[ 1-  (2+\frac{\sqrt{31}}{2})\Phi\overline{\Phi}]$ \\
& $({\bf 1},{\bf 2})_{-\frac{1}{4}}$ &$-1$ & $-1$ & $1$ 
%& $\frac{1}{2}( 3+ \sqrt{19+6\sqrt{31}})$ 
 & $\frac{1}{2}(3\sqrt{31}+5)$ & $T_{\alpha \beta}
\overline{C}_i (1-\Phi \overline{\Phi})^{\frac{\sqrt{31}}{4}-1}
[ 1-  (2+\frac{\sqrt{31}}{2})\Phi\overline{\Phi}]$ \\
& $({\bf 4},{\bf 1})_{+\frac{3}{4}}$ & $3$ &$3$ & $0$
%&$\frac{1}{2}(3+\sqrt{7+2\sqrt{79}})$ 
& $\frac{1}{2}(\sqrt{79}-1)$ 
&  $T_{\alpha \beta}\left[{B}_{(i} {B}_{j)} {B}_k -
\frac{1}{2} \delta_{k(i} {B}_{j)} {B}_l {B}_l \right]
(1-\Phi \overline{\Phi})^{\frac{\sqrt{79}}{4}-1}$  \\
& $({\bf 4},{\bf 1})_{+\frac{1}{4}}$ &$1$ &$1$ & $0$ 
%&$\frac{1}{2}(3+\sqrt{15+2\sqrt{103}})$ 
& 
$\frac{1}{2}(\sqrt{103}+3)$ 
&  $T_{\alpha \beta}\left[{B}_{(i} {B}_{j)} 
\overline{B}_k -\frac{1}{2} \delta_{k(i} {B}_{j)} {B}_l 
\overline{B}_l \right](1-\Phi \overline{\Phi})^{\frac{\sqrt{103}}{4}-1}$  \\
& $({\bf 4},{\bf 1})_{-\frac{1}{4}}$ &$-1$ &$-1$& $0$
%&$\frac{1}{2}(3+\sqrt{15+2\sqrt{103}})$ 
& $\frac{1}{2}(\sqrt{103}+3)$ 
&  $T_{\alpha \beta}\left[{B}_{(i} \overline{B}_{j)} 
\overline{B}_k -\frac{1}{2} \delta_{k(i} {B}_{j)} 
\overline{B}_l \overline{B}_l \right]
(1-\Phi \overline{\Phi})^{\frac{\sqrt{103}}{4}-1}$  \\
& $({\bf 4},{\bf 1})_{-\frac{3}{4}}$ &$-3$ &$-3$ & $0$ 
%&$\frac{1}{2}(3+\sqrt{7+2\sqrt{79}})$ 
& $\frac{1}{2}(\sqrt{79}-1)$ 
&  $T_{\alpha \beta}\left[\overline{B}_{(i}
  \overline{B}_{j)} 
\overline{B}_k -\frac{1}{2} \delta_{k(i} 
\overline{B}_{j)} \overline{B}_l 
\overline{B}_l \right](1-\Phi \overline{\Phi})^{\frac{\sqrt{79}}{4}-1}$  \\
&  $({\bf 3},{\bf 2})_{+\frac{3}{4}}$ &$3$ &$3$& $0$ 
%&$\frac{1}{2}(3+\sqrt{3+2\sqrt{55}})$ 
& $\frac{1}{2}(\sqrt{55}-3)$ & $T_{\alpha \beta} ({B}_i 
{B}_j -\frac{1}{2} \delta_{ij} {B}_l {B}_l) 
{C}_k(1-\Phi \overline{\Phi})^{\frac{\sqrt{55}}{4}-1}$ \\
&  $({\bf 3},{\bf 2})_{+\frac{1}{4}}$ &$1$ & $1$ & $0$ 
%&$\frac{1}{2}(3+\sqrt{11+2\sqrt{79}})$ 
& $\frac{1}{2}(\sqrt{79}+1)$ & $T_{\alpha \beta} ({B}_i 
\overline{B}_j -\frac{1}{2} \delta_{ij} {B}_l \overline{B}_l) 
{C}_k(1-\Phi \overline{\Phi})^{\frac{\sqrt{79}}{4}-1}$ \\
&  $({\bf 3},{\bf 2})_{-\frac{1}{4}}$ & $-1$ & $-1$ & $0$ 
%&$\frac{1}{2}(3+\sqrt{11+2\sqrt{79}})$ 
& $\frac{1}{2}(\sqrt{79}+1)$ & $T_{\alpha \beta} ({B}_i 
\overline{B}_j -\frac{1}{2} \delta_{ij} {B}_l \overline{B}_l) 
\overline{C}_k(1-\Phi \overline{\Phi})^{\frac{\sqrt{79}}{4}-1}$ \\
&  $({\bf 3},{\bf 2})_{+\frac{1}{4}}$ &$1$ & $1$ & $0$ 
%&$\frac{1}{2}(3+\sqrt{11+2\sqrt{79}})$ 
& $\frac{1}{2}(\sqrt{79}+1)$ & $T_{\alpha \beta} ({B}_i 
{B}_j -\frac{1}{2} \delta_{ij} {B}_l {B}_l) 
\overline{C}_k(1-\Phi \overline{\Phi})^{\frac{\sqrt{79}}{4}-1}$ \\
&  $({\bf 3},{\bf 2})_{-\frac{1}{4}}$ &$-1$ & $-1$ & $0$ 
%&$\frac{1}{2}(3+\sqrt{11+2\sqrt{79}})$ 
& $\frac{1}{2}(\sqrt{79}+1)$ & $T_{\alpha \beta}
(\overline{B}_i 
\overline{B}_j -\frac{1}{2} \delta_{ij} 
\overline{B}_l \overline{B}_l) {C}_k(1-\Phi 
\overline{\Phi})^{\frac{\sqrt{79}}{4}-1}$ \\
&  $({\bf 3},{\bf 2})_{-\frac{3}{4}}$ &$-3$ & $-3$ & $0$ 
%&$\frac{1}{2}(3+\sqrt{3+2\sqrt{55}})$ 
& $\frac{1}{2}(\sqrt{55}-3)$ & $T_{\alpha \beta}
(\overline{B}_i 
\overline{B}_j -\frac{1}{2} \delta_{ij} 
\overline{B}_l \overline{B}_l) 
\overline{C}_k(1-\Phi \overline{\Phi})^{\frac{\sqrt{55}}{4}-1}$ \\
&  $({\bf 2},{\bf 3})_{+\frac{3}{4}}$ &$3$ &$3$& $0$ 
%&$\frac{1}{2}(3+\sqrt{3+2\sqrt{55}})$ 
& $\frac{1}{2}(\sqrt{55}-3)$ & $T_{\alpha \beta} ({C}_i 
{C}_j -\frac{1}{2} \delta_{ij} {C}_l {C}_l) 
{B}_k(1-\Phi \overline{\Phi})^{\frac{\sqrt{55}}{4}-1}$ \\
&  $({\bf 2},{\bf 3})_{+\frac{1}{4}}$ &$1$ &$1$& $0$ 
%&$\frac{1}{2}(3+\sqrt{11+2\sqrt{79}})$ 
& $\frac{1}{2}(\sqrt{79}+1)$ & $T_{\alpha \beta} ({C}_i 
\overline{C}_j -\frac{1}{2} \delta_{ij} {C}_l \overline{C}_l) 
{B}_k(1-\Phi \overline{\Phi})^{\frac{\sqrt{79}}{4}-1}$ \\
&  $({\bf 2},{\bf 3})_{-\frac{1}{4}}$ &$-1$ &$-1$& $0$ 
%&$\frac{1}{2}(3+\sqrt{11+2\sqrt{79}})$ 
& $\frac{1}{2}(\sqrt{79}+1)$ & $T_{\alpha \beta} ({C}_i 
\overline{C}_j -\frac{1}{2} \delta_{ij} {C}_l \overline{C}_l) 
\overline{B}_k(1-\Phi \overline{\Phi})^{\frac{\sqrt{79}}{4}-1}$ \\
&  $({\bf 2},{\bf 3})_{+\frac{1}{4}}$ &$1$ &$1$ & $0$ 
%&$\frac{1}{2}(3+\sqrt{11+2\sqrt{79}})$ 
& $\frac{1}{2}(\sqrt{79}+1)$ & $T_{\alpha \beta} ({C}_i 
{C}_j -\frac{1}{2} \delta_{ij} {C}_l {C}_l) 
\overline{B}_k(1-\Phi \overline{\Phi})^{\frac{\sqrt{79}}{4}-1}$ \\
&  $({\bf 2},{\bf 3})_{-\frac{1}{4}}$ &$-1$ & $-1$& $0$ 
%&$\frac{1}{2}(3+\sqrt{11+2\sqrt{79}})$ 
& $\frac{1}{2}(\sqrt{79}+1)$ & $T_{\alpha \beta}
(\overline{C}_i 
\overline{C}_j -\frac{1}{2} \delta_{ij} 
\overline{C}_l \overline{C}_l) {B}_k(1-\Phi 
\overline{\Phi})^{\frac{\sqrt{79}}{4}-1}$ \\
&  $({\bf 2},{\bf 3})_{-\frac{3}{4}}$ &$-3$ & $-3$ & $0$ 
%&$\frac{1}{2}(3+\sqrt{3+2\sqrt{55}})$ 
& $\frac{1}{2}(\sqrt{55}-3)$ & $T_{\alpha \beta}
(\overline{C}_i 
\overline{C}_j -\frac{1}{2} \delta_{ij} 
\overline{C}_l \overline{C}_l) 
\overline{B}_k(1-\Phi \overline{\Phi})^{\frac{\sqrt{55}}{4}-1}$ \\
&  $({\bf 1},{\bf 4})_{+\frac{3}{4}}$ &$3$ & $3$ & $0$ 
%&$\frac{1}{2}(3+\sqrt{7+2\sqrt{79}})$ 
& $\frac{1}{2}(\sqrt{79}-1)$ & $T_{\alpha \beta}
\left[{C}_{(i} {C}_{j)} {C}_k -\frac{1}{2} \delta_{k(i} {C}_{j)} {C}_l 
{C}_l \right](1-\Phi \overline{\Phi})^{\frac{\sqrt{79}}{4}-1}$ \\
&  $({\bf 1},{\bf 4})_{+\frac{1}{4}}$ &$1$ & $1$ & $0$ 
%&$\frac{1}{2}(3+\sqrt{15+2\sqrt{103}})$ 
& $\frac{1}{2}(\sqrt{103}+3)$ & $T_{\alpha \beta}
\left[{C}_{(i} {C}_{j)} \overline{C}_k -\frac{1}{2} \delta_{k(i}
  {C}_{j)} {C}_l 
\overline{C}_l \right](1-\Phi \overline{\Phi})^{\frac{\sqrt{103}}{4}-1}$ \\
&  $({\bf 1},{\bf 4})_{-\frac{1}{4}}$ &$-1$ & $-1$ & $0$ 
%&$\frac{1}{2}(3+\sqrt{15+2\sqrt{103}})$ 
& $\frac{1}{2}(\sqrt{103}+3)$ & $T_{\alpha \beta}
\left[{C}_{(i} \overline{C}_{j)} \overline{C}_k -\frac{1}{2} 
\delta_{k(i} {C}_{j)} 
\overline{C}_l \overline{C}_l \right](1-\Phi 
\overline{\Phi})^{\frac{\sqrt{103}}{4}-1}$ \\
&  $({\bf 1},{\bf 4})_{-\frac{3}{4}}$ &$-3$ & $-3$ & $0$ 
%&$\frac{1}{2}(3+\sqrt{7+2\sqrt{79}})$ 
& $\frac{1}{2}(\sqrt{79}-1)$ & $T_{\alpha \beta}
\left[\overline{C}_{(i} \overline{C}_{j)} \overline{C}_k -\frac{1}{2} \delta_{k(i} 
\overline{C}_{j)} \overline{C}_l 
\overline{C}_l \right](1-\Phi \overline{\Phi})^{\frac{\sqrt{79}}{4}-1}$ \\
\hline
\end{tabular} 
\end{center}
\caption{\sl 
The next level of spin-2 component of the massive graviton multiplets. For each
multiplet we present $SO(8)$ representation (\ref{sobranching}),
$SU(2) \times SU(2) \times U(1)_{R}$ representations
denoted by $({\bf 2l_1+1},{\bf 2l_2+1 })_R$, 
$R_{\phi}$-charge, $R_{\psi}$-charge, the KK
excitation number $j$, the mass-squared $m^2 \hat{L}^2$ (\ref{square}) 
of the $AdS_4$ field
and the corresponding ${\cal N}=2$  SCFT operator.  
The dimension $\Delta$ (\ref{Del}) of the spin-2 component
of the multiplet we do not present here due to 
the space can be obtained also.}
%\label{tableso4}
\end{table} 
%%%%%%%%%%%%%%%%%%%%%%%%%%%%%%%%%%%%%%%%%%%%%%%%%%%%%%%%%%%%%%%%%%%%%%%%%%%%
%%%%%%%%%%%%%%%%%%%%%%%%%%%%%%%%%%%%%%%%%%%%%%%%%%%%%%%%%%%%%%%%%%%%%%%%%%%%

One can extend these procedures to the higher excitations. For example,
at the level of $n=4$ excitation, by extending the procedure in
(\ref{sobranching})
one step further,
one has the following branching rules
\bea
&& {\bf 294}_v(4,0,0,0)  \rightarrow  {\bf 294}(0,0,4) \rightarrow
{\bf 35} \oplus \overline{\bf 35} \oplus {\bf 70} \oplus
\overline{\bf 70} \oplus {\bf 84} 
\label{n4}
\\
&& \rightarrow  
[({\bf 1},{\bf 1})   
\oplus  3 ({\bf 3},{\bf 1}) \oplus  4 ({\bf 2},{\bf 2}) \oplus 3
({\bf 1},{\bf 3})] \oplus  5 ({\bf 5},{\bf 1}) \oplus  8 ({\bf 4},{\bf
  2})
 \oplus  9 ({\bf 3},{\bf 3}) \oplus  8 ({\bf 2},{\bf
  4})  \oplus  5 ({\bf 1},{\bf 5}).
\nonu
\eea
Then one obtains KK excitation $j=2$ state for $({\bf 1},{\bf 1 })$ and
$j=1$ states for $({\bf 3},{\bf 1 }), ({\bf 2},{\bf 2})$ and $({\bf
  1},{\bf 3})$ inside of bracket in (\ref{n4}). 
The remaining five independent new states 
with $j=0$ arise also in this level. 
One can analyze also the corresponding 
${\cal N}=2$ SCFT operators. 
For example, some(three states from nine possible states) 
of the wavefunctions for $l_1=l_2=1$ with $R=0$
corresponding to $({\bf 3},{\bf 3})_0$ are given by
\bea
T_{\alpha \beta}
{B}_1 \overline{B}_2 {C}_1 \overline{C}_2
(1-\Phi \overline{\Phi})^{\sqrt{7}-1}  & \leftrightarrow &
T_{\alpha \beta} e^{i(\phi_1 +\phi_2)} \sin \theta_1 \sin
\theta_2 (1-\Phi \overline{\Phi})^{\sqrt{7}-1}, \nonu \\ 
T_{\alpha \beta}
({B}_1 \overline{B}_1 - {C}_1 \overline{C}_1){C}_1
\overline{C}_2(1-\Phi \overline{\Phi})^{\sqrt{7}-1}  & \leftrightarrow &
T_{\alpha \beta} e^{i\phi_2} \cos \theta_1 \sin
\theta_2 (1-\Phi \overline{\Phi})^{\sqrt{7}-1}, \nonu \\ 
T_{\alpha \beta}
{B}_2 \overline{B}_1 {C}_2 \overline{C}_1 
(1-\Phi \overline{\Phi})^{\sqrt{7}-1} 
& \leftrightarrow &
T_{\alpha \beta} e^{-i(\phi_1 +\phi_2)} \sin \theta_1 \sin
\theta_2 (1-\Phi \overline{\Phi})^{\sqrt{7}-1}
\label{states}
\eea
where the corresponding $(m_1, m_2)$ values are $(1,1), (0,1)$ and
$(-1,-1)$ 
respectively \cite{KM}.
One can easily see these from (\ref{JJs}) or (\ref{fivefields}). 
For the first case of (\ref{states}), 
we take $j_1(\theta_1)$ which is equal to $\sin \theta_1$ and 
$j_1(\theta_2)$ which is equal to $\sin \theta_2$, as an eigenfunctions. 
For the second case, 
we take $j_1(\theta_1)$ which is equal to $\cos \theta_1$ and 
$j_1(\theta_2)$ that is equal to $\sin \theta_2$.
For the last, we take  $j_2(\theta_1)$ that is $\frac{1}{2} \sin
\theta_1$ and 
$j_2(\theta_2)$ that is $\frac{1}{2} \sin \theta_2$, 
as the regular solutions. Note that the regularity of the solutions
(\ref{JJs}) depend on the magnitude of $(m_1,m_2)$
with respect to the $R$ charge.
In general, one expects that for general quantum numbers $l_1, l_2$
and $R$ of
$SU(2) \times SU(2) \times U(1)_R$, the operator is given by the product of   
$T_{\alpha \beta}$ with several ${B}_i$ or
${C}_j$'s(and its conjugated fields) and some function of $\Phi\overline{\Phi}$.
For nonzero $j$'s, the explicit form 
of hypergeometric functions is
needed to identify the corresponding ${\cal N}=2$ SCFT operators and 
there exists a 
polynomial up to the order $j$ in $\Phi\overline{\Phi}$.

%%%%%%%%%%%%%%%%%%%%%%%%%%%%%%%%%%%%%%%%%%%%%%%%%%%%%%%%%%%%%%%%%%%%%%%%%%%%%%%
%%%%%%%%%%%%%%%%%%%%%%%%%%%%%%%%%%%%%%%%%%%%%%%%%%%%%%%%%%%%%%%%%%%%%%%%%%%%%%%%
\section{
Conclusions and outlook }
%%%%%%%%%%%%%%%%%%%%%%%%%%%%%%%%%%%%%%%%%%%%%%%%%%%%%%%%%%%%%%%%%%%%%%%%%%%%%%%%
%%%%%%%%%%%%%%%%%%%%%%%%%%%%%%%%%%%%%%%%%%%%%%%%%%%%%%%%%%%%%%%%%%%%%%%%%%%%%%%%

We have elaborated the 11-dimensional background geometry originated
from \cite{CPW}. In particular, we have presented the correct
expression for 3-form potential and some relations for $SU(3) \times
U(1)_R$ symmetry are also given. 
We have made the relations between the fields of $AdS_4$ supergravity
and composite operators of the IR boundary gauge theory in Tables 1, 2
and 3.
We have computed the KK reduction for spin-2 excitations around the warped
11-dimensional theory background that is dual to the ${\cal N}=2$
mass-deformed Chern-Simons matter theory with $SU(2) \times SU(2)
\times U(1)_R$ symmetry. 
The spectrum of spin-2 excitations was given by solving the equations
of motion for minimally coupled scalar theory in this background.
The $AdS_4$ mass formula of the KK modes is given by (\ref{square}). 
The quantum number $l_1, l_2$ and $R$ for $SU(2)
\times SU(2) \times U(1)_R$
representation and the KK excitation number $j$ arise in this mass formula.      
We calculated the dimensions of the dual operators in the boundary
${\cal N}=2$ 
SCFT via AdS/CFT correspondence and in Tables 4 and 5 we presented the
summary of this work.

As observed in \cite{CPW},  at $\mu =\frac{\pi}{2}$, the metric has
conical singularity from (\ref{cond}), (\ref{zis}) and (\ref{shift}).
That is, the apex or node is a double point, i.e., a singularity for
which $C=0$ and $d C =0$ where $C$ is a complex manifold (\ref{cond})
but for which the matrix of second derivatives is nondegenerate. 
Then the ${\bf S}^7$ degenerate to the conifold times ${\bf S}^1$.
The solution by the metric $T^{1,1}$ can be related, via T-duality, to the  
Romans' type IIB supergravity theory in 10-dimensions \cite{Romans}.
Furthermore, this leads to the nontrivial Klebanov-Witten fixed point of the
holographic RG flow in \cite{KW}. This is related to 
the fact that there exists a flow between a space that is locally
$AdS_5 \times T^{1,1}$ and the $AdS_4 \times \frac{{\bf S}^7}{{\bf
Z}_k}$ geometry, mentioned in \cite{ABJM}. 
By adding a flux to this solution
and squashing and stretching ellipsoidally, the theory can flow to other nontrivial
fixed point. One of the candidates is given by Pilch-Warner fixed
point in \cite{Pilch}. The gravity solution interpolating these two
fixed points has been constructed in \cite{HPRW}. 
It would be interesting to elaborate on these issues.

The quadratic equation (\ref{cond}) is so-called $A_1$ conifold.
It is an open problem to generalize  this singularity to 
ADE type singularities discussed in \cite{GNS}.
For this approach, one needs to use the Einstein-Kahler metric on 
del Pezzo surface $dP_k, k\geq 3$. According to \cite{FHPR}, 
${\cal C}(T^{1,1}) \times {\bf C}$ can be obtained from either
${\frac{{\bf C}^3}{{\bf Z}_2 \times {\bf Z}_2} \times {\bf C}}$ or 
${\cal C}(dP_3) \times {\bf C}$, in which their superpotentials are the
same for abelian theory and their quiver diagrams looks similar to
each other, by partial resolutions. It would be interesting to find
out the corresponding gravity duals.

In the context of 4-dimensional gauged supergravity, it is known that
very few of critical points(supersymmetric or nonsupersymmetric) 
are found although there are 
70 scalar fields. The lesson we have learned from \cite{CPW} is when
we go to 11-dimensional theory, maybe we will find various
11-dimensional solutions even if the 4-dimensional flow equations are
the same. For given 7-dimensional Sasaki-Einstein spaces, one can
think of the possibilities to have resolved cone over these spaces     
and in the gravity side, the correct 11-dimensional metric should be
found with appropriate field strengths. We expect that since the flow
equations 
in 4-dimensions are related to the ${\cal N}=2$ supersymmetry with $U(1)_R$
charge, other types of 
11-dimensional solutions with common 4-dimensional flow equations will
arise. 

One possibility is characterized by the $SU(2) \times U(1)
\times U(1)_R$ symmetry which is smaller than $SU(2) \times SU(2)
\times U(1)_R$. The symmetry breaking of $SU(2) \times U(1)$ can 
occur from either the $SU(3) \times U(1)_R$ symmetry or $SU(2) \times SU(2)
\times U(1)_R$ symmetry. In this case, the metric corresponding to ${\bf S}^2
\times {\bf S}^2$ should preserve only one ${\bf S}^2$ symmetry.

Another possibility is given by the
$SU(3) \times U(1) \times U(1)_R$ symmetry which is larger than 
$SU(3) \times U(1)_R$ symmetry and can be obtained from 
the symmetry breaking of $SU(4) \times U(1)_R$. Since there exists 
an extra $U(1)$ factor, it is nontrivial to find out the correct
4-form fields that preserve the whole $SU(3) \times U(1) \times
U(1)_R$ symmetry.  
Note that the 4-forms in \cite{CPW}, that are given by (\ref{F4su3}), 
breaks the $SU(3) \times U(1)
\times U(1)_R$ symmetry group to $SU(3) \times U(1)_R$.
It would be interesting to study these issues more detail.
 
\vspace{.7cm}

%%%%%%%%%%%%%%%%%%%%%%%%%%%%%%%%%%%%%%%%%%%%%%%%%%%%%%%%%%%%%%
%%%%%%%%%%%%%%%%%%%%%%%%%%%%%%%%%%%%%%%%%%%%%%%%%%%%%%%%%%%%%%%
\centerline{\bf Acknowledgments}
%%%%%%%%%%%%%%%%%%%%%%%%%%%%%%%%%%%%%%%%%%%%%%%%%%%%%%%%%%%%%%%
%%%%%%%%%%%%%%%%%%%%%%%%%%%%%%%%%%%%%%%%%%%%%%%%%%%%%%%%%%%%%%%

We would like to thank 
S. Franco, Sangmin Lee and F.D. Rocha 
%for mathematica computations and
%O. Lunin  and D. Kutasov 
for discussions. 
%and K. Skenderis for discussions on the 
%domain wall solutions.
This work was supported by the 
National Research Foundation of Korea(NRF) grant 
funded by the Korea government(MEST)(No. 2009-0084601).
%This work was supported by grant No.
%R01-2006-000-10965-0 from the Basic Research Program of the Korea
%Science \& Engineering Foundation.  
%I acknowledge warm hospitality of 
%Particle Theory Group, Enrico Fermi Institute 
%at University of Chicago
%where this work was initiated.

\appendix

\renewcommand{\thesection}{\large \bf \mbox{Appendix~}\Alph{section}}
\renewcommand{\theequation}{\Alph{section}\mbox{.}\arabic{equation}}

%%%%%%%%%%%%%%%%%%%%%%%%%%%%%%%%%%%%%%%%%%%%%%%%%%%%%%%%%%%%%%%%%%%%%%%%
%%%%%%%%%%%%%%%%%%%%%%%%%%%%%%%%%%%%%%%%%%%%%%%%%%%%%%%%%%%%%%%%%%%%%%%%%
\section{The 4-form field strength and the Ricci tensor 
in frame basis for $SU(2) \times SU(2) \times
U(1)_R$-invariant case}
%%%%%%%%%%%%%%%%%%%%%%%%%%%%%%%%%%%%%%%%%%%%%%%%%%%%%%%%%%%%%%%%%%%%%%
%%%%%%%%%%%%%%%%%%%%%%%%%%%%%%%%%%%%%%%%%%%%%%%%%%%%%%%%%%%%%%%%%%%%%%

The set of frames for the 11-dimensional metric (\ref{11dmetric})
is given by 
\bea
e^1  & = & -\frac{1}{\sqrt{\Delta}} \, e^A \, d x^1,
\qquad
e^2   =  \frac{1}{\sqrt{\Delta}} \, e^A \, d x^2,
\qquad
e^3   =  \frac{1}{\sqrt{\Delta}} \, e^A \, d x^3,
\qquad
e^4    =   \frac{1}{\sqrt{\Delta}}  \, d r,
\nonu \\
e^5   & = &  3^{\frac{3}{2}} \,\hat{L}^2 \,
\Delta^{\frac{1}{4}} \, \frac{\sqrt{X}}{\rho^3} \, d \mu,
\nonu \\
 e^6  &  = &  3^{\frac{3}{2}}\, \hat{L}^2\,
\Delta^{\frac{1}{4}} \,\rho \cos \mu \, \frac{1}{\sqrt{6}} \, d \theta_1,
\nonu \\
 e^7 & = & 3^{\frac{3}{2}}\, \hat{L}^2\,
\Delta^{\frac{1}{4}} \,\rho \cos \mu \, \frac{1}{\sqrt{6}} \, \sin \theta_1
\, d \phi_1,
\nonu \\
 e^8  & = &  3^{\frac{3}{2}}\, \hat{L}^2\,
\Delta^{\frac{1}{4}} \,\rho \cos \mu \, \frac{1}{\sqrt{6}} \, d \theta_2,
\nonu \\
 e^9 & = & 3^{\frac{3}{2}}\, \hat{L}^2\,
\Delta^{\frac{1}{4}} \,\rho \cos \mu \, \frac{1}{\sqrt{6}} \, \sin \theta_2
\, d \phi_2,
\nonu \\
 e^{10} & = & 3^{\frac{3}{2}}\, \hat{L}^2\,
\Delta^{\frac{1}{4}} \, \frac{\rho^{5}}{2\sqrt{X}} \, \sin 2\mu \, 
\left[ - \frac{d \psi}{\rho^8}
  + \frac{1}{3}(d \psi + d \phi + \cos \theta_1 \, d \phi_1 + 
\cos \theta_2 \, d \phi_2)
 \right],
\label{11frames}
\\
 e^{11} & = & 3^{\frac{3}{2}}\, \hat{L}^2\,
\Delta^{\frac{1}{4}} \,  \frac{\rho \, \cosh \chi}{\sqrt{X}} \, \left[ \sin^2\mu \, d \psi + 
\frac{1}{3} \cos^2 \mu (d \psi + d \phi + \cos \theta_1 \, d \phi_1 + 
\cos \theta_2 \, d \phi_2)\right],
\nonu
\eea
where 
 \footnote{
The general formula for $
C^{(3)}$ used in \cite{KPR} satisfying the whole $SU(3) \times
U(1)_R$-invariant 
flow equation is
given by
\bea
C^{(3)} =\frac{3^{\frac{9}{4}} \,\hat{L}^3 \,\tanh \chi}{4X} \left(3
z^{[1} \, d z^2 \wedge d z^{3]} \wedge d \overline{w}-\rho^8
\,\overline{w} \, d z^1 \wedge d z^2 \wedge d z^3\right)
\label{C3FKR}
\eea
which is a complex conjugate of $C^{(3)}_{CPW}$ in \cite{CPW} and the
definitions for the complex coordinates $z_i$ and $w$ are given in \cite{KPR}. 
This also occurs in the last equation of section 4 of 
\cite{CPW} which looks similar to (\ref{C3FKR}). Note that there exist
some differences in the overall coefficient and the $\rho$ dependence with 
$3^{\frac{9}{4}} \,\hat{L}^3 = L_{CPW}^3$.
One can check (\ref{C3FKR}) from the ``corrected''(there should be 
a plus sign in $e^{10}$) 3-form given in
\cite{CPW}
by changing the rectangular coordinates to the angular ones.
If we
substitute (\ref{rhochi}), the criticality condition at the IR, 
into (\ref{C3FKR}), then we obtain
$C^{(3)} =\frac{3^{\frac{11}{4}} \,\hat{L}^3}{4(|z_i|^2 + 3 |w|^2)}(
z^{[1} \, d z^2 \wedge d z^{3]} \wedge d \overline{w}-
\,\overline{w} \, d z^1 \wedge d z^2 \wedge d z^3)$ which is the same
as (12) of \cite{KPR}. }
we use (\ref{X}), (\ref{delta}) and (\ref{rhochi}) with
the $AdS_4$ radius $\hat{L} \equiv 3^{-\frac{3}{4}} L$.  

It turns out that the antisymmetric 
field strengths have the following nonzero components
in the orthonormal frame basis used in (\ref{c3}) or in
(\ref{11frames}) \footnote{Let us mention that although these 4-forms
satisfy the whole RG flows, we are interested in the critical values 
(\ref{rhochi}), then the supergravity fields dependence does not
appear. In principle, one can write down the general 4-forms with
$(\rho,\chi)$ dependence from (\ref{c3}) and (\ref{a3}) 
explicitly where one should use the RG flow
equations \cite{AP}. The corresponding Ricci tensor components with
$(\rho,\chi)$
dependence can be obtained from (\ref{11dmetric}) 
or (\ref{11frames}).}
\bea
F_{1234}  & = & -\frac{3 \cdot  2^{\frac{1}{3}} \cdot 3^{\frac{3}{4}}}
{\hat{L}(2-\cos2\mu)^{\frac{4}{3}}}, \qquad
F_{568\,10} = \frac{2^{\frac{1}{3}} \cdot 3^{\frac{3}{4}}
  \sin(\phi+2\psi) \sin2\mu}
{\hat{L}(2-\cos2\mu)^{\frac{4}{3}}}=-F_{579\,10}, 
\nonu \\
F_{568\,11} & = & -\frac{2^{\frac{5}{6}} \cdot 3^{\frac{3}{4}} \sin(\phi+2\psi)}
{\hat{L}(2-\cos
  2\mu)^{\frac{1}{3}}}=-F_{579\,11}=-F_{69\,10\,11}=-F_{78\,10\,11}, 
\nonu \\
F_{569\,10}  & = &  -\frac{2^{\frac{1}{3}} \cdot 3^{\frac{3}{4}} 
\cos(\phi+2\psi) \sin2\mu}
{\hat{L}(2-\cos 2\mu)^{\frac{4}{3}}}=F_{578\,10},
\nonu \\
F_{569\,11} & = & \frac{2^{\frac{5}{6}} \cdot 3^{\frac{3}{4}} \cos(\phi+2\psi)}
{\hat{L}(2-\cos
  2\mu)^{\frac{1}{3}}}=F_{578\,11}=F_{68\,10\,11}=-F_{79\,10\,11}, 
\label{F4}
\eea
where the angle-dependences for $\phi$ and $\psi$ appear in the
combination of $(\phi + 2\psi)$. One can make the two $U(1)$ symmetries 
generated by $\phi$ and $\psi$ 
which preserve this combination $(\phi +2 \psi)$.
After substituting (\ref{F4}) into 
the right hand side of Einstein equation (\ref{fieldequations})
with frame basis (\ref{11frames}) 
one reproduces the one of $SU(3) \times U(1)_R$ case
\cite{CPW,KPR} exactly. 
On the other hand, the Ricci tensor in the
frame basis (\ref{11frames}) is identical to the one with $SU(3)
\times U(1)_R$ symmetry in \cite{CPW}. 
Therefore, one concludes that the solutions (\ref{F4}) indeed 
satisfy the 11-dimensional Einstein-Maxwell equations.
Let us present the Ricci tensor components with frame basis here for convenience.
\bea
R_1^{\, 1} & = & -\frac{(55-32 \cos 2\mu + 3 \cos 4\mu) }
{3 \cdot 2^{\frac{1}{3}} \,\sqrt{3} \, \hat{L}^2\,
  (2-\cos 2\mu )^{\frac{8}{3}}} =R_2^{\, 2} =R_3^{\, 3}=R_4^{\, 4}
= -2 R_6^{\, 6} = -2 R_7^{\, 7} = -2 R_8^{\, 8}=-2 R_9^{\, 9},
\nonu \\
R_5^{\, 5} & = & \frac{(29-16 \cos 2\mu) }
{3 \cdot 2^{\frac{1}{3}} \,\sqrt{3} \,\hat{L}^2\,
  (2-\cos 2\mu )^{\frac{8}{3}}} = R_{10}^{\, 10}, \qquad
R_{10}^{\, 11} =  -\frac{2 \cdot 2^{\frac{1}{6}} \sin 2\mu }
{\sqrt{3} \,\hat{L}^2\,
  (2-\cos 2\mu )^{\frac{5}{3}}} = R_{11}^{\,\,10},
\nonu \\
R_{11}^{\, 11} & = & \frac{(80-64 \cos 2\mu +9 \cos 4 \mu) }
{3 \cdot 2^{\frac{1}{3}} \,\sqrt{3} \,\hat{L}^2\,
  (2-\cos 2\mu )^{\frac{8}{3}}},
\label{Ricci}
\eea
and other components vanish.
This implies that  the Einstein-Maxwell 
equations (\ref{fieldequations}) with frame basis are satisfied \footnote{ 
Also let us mention that the 3-form potential appeared
between (4.28) and (4.29) in \cite{CPW} has an error in the sign of $e^{10}$.
See also (\ref{3cpw}).
After this correction, the 3-form of \cite{CPW} is exactly the complex
conjugation of 3-form in \cite{KPR}. Since the full 3-form has its
complex conjugation, eventually the 4-form \cite{CPW} is the same as
the one in \cite{KPR}.}. 
In Appendix B, we write down the Ricci tensor and 4-forms in the
coordinate basis and in Appendix C, we present the 4-forms in the
frame basis for $SU(3) \times U(1)_R$-invariant case.

%%%%%%%%%%%%%%%%%%%%%%%%%%%%%%%%%%%%%%%%%%%%%%%%%%%%%%%%%%%%%%%%%%%%%%%%
%%%%%%%%%%%%%%%%%%%%%%%%%%%%%%%%%%%%%%%%%%%%%%%%%%%%%%%%%%%%%%%%%%%%%%%%%
\section{The Ricci tensor and the 4-form 
field strength in coordinate basis for $SU(2) \times SU(2) \times
U(1)_R$-invariant case}
%%%%%%%%%%%%%%%%%%%%%%%%%%%%%%%%%%%%%%%%%%%%%%%%%%%%%%%%%%%%%%%%%%%%%%
%%%%%%%%%%%%%%%%%%%%%%%%%%%%%%%%%%%%%%%%%%%%%%%%%%%%%%%%%%%%%%%%%%%%%%

The $SU(2) \times SU(2) \times U(1)_R$-invariant 11-dimensional metric 
(\ref{11dmetric}) explained in section 2 is given by
\bea
ds_{11}^2 =\Delta^{-1}\left(dr^2 +e^{2 A(r)}
\eta_{\mu\nu}dx^\mu dx^\nu \right)+3^{\frac{3}{2}} \hat{L}^2
\sqrt{\Delta} \,
ds_7^2(\rho,\chi).
\label{Met}
\eea
This (\ref{Met}) generates the Ricci tensor components in the
coordinate basis:
\bea
\hat{R}_1^{\, 1} & = & \frac{(-252 +241 c_{2\mu} -44 c_{4\mu} + 3 c_{6\mu}) }
{6 \cdot 2^{\frac{1}{3}} \,\sqrt{3} \, \hat{L}^2\,
  (2-c_{2\mu} )^{\frac{11}{3}}} =\hat{R}_2^{\, 2} =\hat{R}_3^{\, 3}
=\hat{R}_4^{\, 4} =
-\frac{1}{2} \hat{R}_6^{\, 6}=-\frac{1}{2} \hat{R}_8^{\, 8},
\nonu \\
\hat{R}_5^{\, 5} & = & \frac{(29-16 c_{2\mu}) }
{3 \cdot 2^{\frac{1}{3}} \,\sqrt{3} \,\hat{L}^2\,
  (2-c_{2\mu} )^{\frac{8}{3}}}, 
\qquad
\hat{R}_7^{\, 7}   =   \frac{(55 -32 c_{2\mu} +3 c_{4\mu})  }
{6 \cdot 2^{\frac{1}{3}}\, \sqrt{3} \,\hat{L}^2\,
  (2-c_{2\mu} )^{\frac{8}{3}}}=\hat{R}_{9}^{\, 9},
\nonu \\
\hat{R}_{7}^{\, 10} & = & \frac{ c_{\theta_1}\, c^2_{\mu}\, 
(21-14 c_{2\mu} + c_{4\mu}) }
{2^{\frac{1}{3}} \,\sqrt{3} \,\hat{L}^2\,
  (2-c_{2\mu} )^{\frac{14}{3}}},
\qquad
\hat{R}_{7}^{\, 11}   =   \frac{c_{\theta_1}\,(-3 +c_{2\mu}) \,
(-3 c_{\mu} + c_{3\mu})^2 }
{2 \cdot 2^{\frac{1}{3}} \,\sqrt{3} \,\hat{L}^2\,
  (2-c_{2\mu} )^{\frac{14}{3}}},
\nonu \\
\hat{R}_{9}^{\, 10} & = & \frac{c_{\theta_2}\, c^2_{\mu} \,
(21-14 c_{2\mu} + c_{4\mu}) }
{2^{\frac{1}{3}} \,\sqrt{3} \,\hat{L}^2\,
  (2-c_{2\mu} )^{\frac{14}{3}}},
\qquad
\hat{R}_{9}^{\, 11}   =   \frac{c_{\theta_2}\,(-3 +c_{2\mu}) \,
(-3 c_{\mu} + c_{3\mu})^2 }
{2 \cdot 2^{\frac{1}{3}} \,\sqrt{3} \,\hat{L}^2\,
  (2-c_{2\mu} )^{\frac{14}{3}}},
\nonu \\
\hat{R}_{10}^{\, 10} & = & -\frac{(207+208 c_{2\mu} 
-41 c_{4\mu} +3 c_{6\mu}) }
{3 \cdot 2^{\frac{1}{3}} \,\sqrt{3} \,\hat{L}^2\,
  (2-c_{2\mu} )^{\frac{11}{3}}},
\qquad
\hat{R}_{10}^{\, 11}   =   -\frac{4 }
{2^{\frac{1}{3}} \,\sqrt{3} \,\hat{L}^2\,
  (2-c_{2\mu} )^{\frac{2}{3}}},
\nonu \\
\hat{R}_{11}^{\, 10} & = &  \frac{c^2_{\mu}\, 
(21-14 c_{2\mu} + c_{4\mu}) }
{2^{\frac{1}{3}} \,\sqrt{3} \,\hat{L}^2\,
  (2-c_{2\mu} )^{\frac{14}{3}}},
\qquad
\hat{R}_{11}^{\, 11}  =  \frac{(20-22 c_{2\mu} + 3c_{4\mu}) }
{3 \cdot 2^{\frac{1}{3}} \,\sqrt{3} \,\hat{L}^2\,
  (2-c_{2\mu} )^{\frac{8}{3}}}.
\label{Riccicoor}
\eea
Or from the orthonormal basis (\ref{11frames}) where $e^a = e^a_m \, d
y^m$ 
and the Ricci tensor
components (\ref{Ricci}) with that basis, 
one obtains (\ref{Riccicoor}) via $\hat{R}_{mn} = e_m^a \, e_n^b \, R_{ab}$.
In this basis, the Ricci tensor also depends on the angular coordinates
$\theta_1$ and $\theta_2$. 
From the (\ref{F4}) in the frame basis, one also writes down them in
the coordinate basis using $\hat{F}_{mnpq} = e_m^a \, e_n^b \, e_p^c
\, e_q^d \, 
F_{abcd}$ as follows: 
\bea
\hat{F}_{1234}  & = & -\frac{9 \cdot  3^{\frac{3}{4}} 
e^{\frac{3r}{\hat{L}}}}
{2\hat{L}}, 
\qquad
\hat{F}_{5678}  =  
\frac{3 \cdot 3^{\frac{3}{4}}\,\hat{L}^3 \,c_{\theta_1} 
\,c^4_{\mu} \,s_{\phi+2\psi}}
{(2-c_{2\mu})^{2}}, 
\nonu \\
\hat{F}_{5679} & = & -\frac{3 \cdot 3^{\frac{3}{4}}\,\hat{L}^3
\,c_{\theta_1} \,s_{\theta_2}
\,c^4_{\mu} \,c_{\phi+2\psi}}
{(2-c_{2\mu})^{2}}, 
\qquad
\hat{F}_{5689}  =   -\frac{3 \cdot 3^{\frac{3}{4}}\,\hat{L}^3 
\,c_{\theta_2}
\,c^4_{\mu} \,s_{\phi+2\psi}}
{(2-c_{2\mu})^{2}}, 
\nonu \\
\hat{F}_{568\,10} & = & -3 \cdot 3^{\frac{3}{4}}\,\hat{L}^3
 \, c^2_{\mu} \,s_{\phi+2\psi}, 
\qquad
\hat{F}_{568\,11}  =  -\frac{3 \cdot 3^{\frac{3}{4}}
\,\hat{L}^3 \,c^4_{\mu} \,s_{\phi+2\psi}}
{(2-c_{2\mu})^{2}},
\nonu \\ 
\hat{F}_{569\,10} &  = & 3 \cdot 3^{\frac{3}{4}}\,\hat{L}^3\,
  c^2_{\mu} \,s_{\theta_2} \,c_{\phi+2\psi},
\qquad
\hat{F}_{569\,11}  =  \frac{3 \cdot 3^{\frac{3}{4}}\,
\hat{L}^3 \,c^4_{\mu}
  \,s_{\theta_2} \,c_{\phi+2\psi}}
{(2-c_{2\mu})^{2}},
\nonu \\ 
\hat{F}_{5789}  & = &  \frac{3 \cdot 3^{\frac{3}{4}}\,\hat{L}^3 \,c^4_{\mu}
  \,c_{\theta_2}\,s_{\theta_1} \,c_{\phi+2\psi}}
{(2-c_{2\mu})^{2}},
\qquad
\hat{F}_{578\,10}  =   3 \cdot 3^{\frac{3}{4}}\,\hat{L}^3\,
  \,c^2_{\mu} \,s_{\theta_1} \,c_{\phi+2\psi}, 
\nonu \\
\hat{F}_{578\,11}  & = &  \frac{3 \cdot 3^{\frac{3}{4}}\,
\hat{L}^3 \,c^4_{\mu}
  \,s_{\theta_1} \,c_{\phi+2\psi}}
{(2-c_{2\mu})^{2}},
\qquad
\hat{F}_{579\,10}  =   3 \cdot 3^{\frac{3}{4}}\,\hat{L}^3\,
  c^2_{\mu} \,s_{\theta_1} \,s_{\theta_2} \,s_{\phi+2\psi}, 
\nonu \\
\hat{F}_{579\,11}  & = &  \frac{3 \cdot 3^{\frac{3}{4}}\,
\hat{L}^3 \,c^4_{\mu}
  \,s_{\theta_1} \,s_{\theta_2} \,s_{\phi+2\psi}}
{(2-c_{2\mu})^{2}},
\qquad
\hat{F}_{678\,10}  =   \frac{3 \cdot 3^{\frac{3}{4}}\,\hat{L}^3
  \,c^3_{\mu}\,s_{\mu} \,c_{\theta_1} \,c_{\phi+2\psi}}{(-2+c_{2\mu})}, 
\nonu \\
\hat{F}_{679\,10}  & = &  \frac{3 \cdot 3^{\frac{3}{4}}\,\hat{L}^3 
\,c^3_{\mu}\,s_{\mu}
  \,c_{\theta_1} \,s_{\theta_2} \,s_{\phi+2\psi}}
{(-2+c_{2\mu})},
\qquad
\hat{F}_{689\,10}  =   -\frac{3 \cdot 3^{\frac{3}{4}}\,\hat{L}^3\,
  \,c^3_{\mu}\,s_{\mu} \,c_{\theta_2}\,c_{\phi+2\psi}}{(-2+c_{2\mu})}, 
\nonu \\
\hat{F}_{68\,10\,11}  & = &   \frac{3 \cdot 3^{\frac{3}{4}}\,
\hat{L}^3 \,c^3_{\mu}\,s_{\mu}
   \,c_{\phi+2\psi}}
{(-2+c_{2\mu})},
\qquad
\hat{F}_{69\,10\,11}   =   \frac{3 \cdot 3^{\frac{3}{4}}\,
\hat{L}^3 \,c^3_{\mu}\,s_{\mu}
   \,s_{\theta_2} \,s_{\phi+2\psi}}
{(-2+c_{2\mu})}, \nonu \\
\hat{F}_{789\,10} & = &  -\frac{3 \cdot 3^{\frac{3}{4}}\,\hat{L}^3\,
  c^3_{\mu}\,s_{\mu} \,s_{\theta_1} \,c_{\theta_2}\,
s_{\phi+2\psi}}{(-2+c_{2\mu})}, 
\qquad
\hat{F}_{78\,10\,11}   =    \frac{3 \cdot 3^{\frac{3}{4}}\,\hat{L}^3
  \,c^3_{\mu}\,s_{\mu} \,s_{\theta_1} \,
s_{\phi+2\psi}}{(-2+c_{2\mu})},
\nonu \\
\hat{F}_{79\,10\,11}  & = &    -\frac{3 
\cdot 3^{\frac{3}{4}}\,\hat{L}^3
  \,c^3_{\mu}\,s_{\mu} \,s_{\theta_1} \,s_{\theta_2} 
\,c_{\phi+2\psi}}{(-2+c_{2\mu})}.
\label{F44}
\eea
In this basis, the 4-form fields depend on the angular coordinates
$\theta_1$ and $\theta_2$ as well as $\mu, \phi$ and $\psi$. 
One sees that the 4-forms in (\ref{F44}) 
contain the combination of $(\phi+2\psi)$ indicating that these
4-forms preserve $U(1)_R$ charge: $\phi \rightarrow \phi + 2\gamma$
and $\psi \rightarrow \psi -\gamma$ explained in (\ref{R}).

%%%%%%%%%%%%%%%%%%%%%%%%%%%%%%%%%%%%%%%%%%%%%%%%%%%%%%%%%%%%%%%%%%%%%
%%%%%%%%%%%%%%%%%%%%%%%%%%%%%%%%%%%%%%%%%%%%%%%%%%%%%%%%%%%%%%%%%%%%%
\section{The 4-form 
field strength in frame basis for $SU(3) \times U(1)_R$-invariant case}
%%%%%%%%%%%%%%%%%%%%%%%%%%%%%%%%%%%%%%%%%%%%%%%%%%%%%%%%%%%%%%%%%%%%
%%%%%%%%%%%%%%%%%%%%%%%%%%%%%%%%%%%%%%%%%%%%%%%%%%%%%%%%%%%%%%%%%%%%

Let us present the 4-form fields in the frame basis used in \cite{CPW}:
\bea
F_{1234}  & = & -\frac{3 \cdot  2^{\frac{1}{3}} \cdot 3^{\frac{3}{4}}}
{\hat{L}(2-c_{2\mu})^{\frac{4}{3}}}, \qquad
F_{567\,10} = \frac{2^{\frac{1}{3}} \cdot 3^{\frac{3}{4}}
  \,s_{3\phi+4\psi}\,(-3c_{\mu}+c_{3\mu})\, s_{\mu}}
{\hat{L}\,(2-c_{2\mu})^{\frac{7}{3}}}=F_{589\,10}, 
\nonu \\
F_{567\,11} & = & \frac{2^{\frac{5}{6}} \cdot 3^{\frac{3}{4}} s_{3\phi+4\psi}}
{\hat{L}\,(2-c_{2\mu})^{\frac{1}{3}}}=F_{589\,11}=F_{68\,10\,11}, 
\qquad
F_{568\,10}   =   \frac{2^{\frac{1}{3}} \cdot 3^{\frac{3}{4}}
  \,c_{3\phi+4\psi}\,(-3c_{\mu}+c_{3\mu})\, s_{\mu}}
{\hat{L}\,(2-c_{2\mu})^{\frac{7}{3}}}, 
\nonu \\
F_{568\,11}  & = &  \frac{  2^{\frac{5}{6}} \cdot 3^{\frac{3}{4}} \,c_{3\phi+4\psi}}
{\hat{L}(2-c_{2\mu})^{\frac{1}{3}}},
\qquad
F_{579\,10}  =  -\frac{2^{\frac{1}{3}} \cdot 3^{\frac{3}{4}} 
\,c_{3\phi+4\psi}\,(-3c_{\mu}+c_{3\mu})\, s_{\mu} }
{\hat{L}\,(2-c_{2\mu})^{\frac{7}{3}}}, 
\nonu \\
F_{579\,11} & = &  -\frac{2^{\frac{5}{6}} \cdot 3^{\frac{3}{4}}
  \,c_{3\phi+4\psi}}
{\hat{L}\,(2-c_{2\mu})^{\frac{1}{3}}}
=F_{67\,10\,11}=F_{89\,10\,11},
\qquad
 F_{79\,10\,11}  =   -\frac{2^{\frac{5}{6}} \cdot 3^{\frac{3}{4}}
  \,s_{3\phi+4\psi}}
{\hat{L}\,(2-c_{2\mu})^{\frac{1}{3}}}.
\label{F4su3}
\eea
This can be obtained from the 3-form potential 
\bea
C^{(3)} = \frac{1}{4} \sinh \chi \, e^{i(-3\phi-4\psi)} \, (e^5+ i e^{10}) 
\wedge (e^6 - i e^9) \wedge (e^7 - i e^8),
\label{3cpw}
\eea
and (\ref{a3})
where the orthonormal frame basis used in (\ref{3cpw}) is given in \cite{CPW}.
In (\ref{F4su3}), the angle-dependences on $\phi$ and $\psi$ arise in
the combination of $(3\phi+4\psi)$ where $\phi \rightarrow \frac{4}{3}
\gamma$ and $\psi \rightarrow \psi -\gamma$ and these disappear by squaring
4-forms 
in the right hand side of 11-dimensional Einstein equation 
(\ref{fieldequations}). This is consistent with the fact that the left
hand side of Einstein equation is given by (\ref{Ricci}) which do not
depend on these angular variables.
One obtains also  the corresponding 4-forms and Ricci tensor
components in the coordinate basis using $\hat{R}_{mn} = e_m^a \, e_n^b \,
R_{ab}$ and $\hat{F}_{mnpq} = e_m^a \, e_n^b \, e_p^c
\, e_q^d \, F_{abcd}$ as mentioned before from (\ref{F4su3}) and (\ref{Ricci}). 
It turns out that they depend on the angular variables $\alpha_1$ and $\theta$.
In particular, the 4-forms contain the trignometric functions with the
argument $(\alpha_3 + 3\phi + 4\psi)$.
By substituting these into the 11-dimensional Einstein equation, one
sees that the dependence on $(\alpha_3 + 3\phi + 4\psi)$ completely
disappears in the right hand side of equation.
This is also consistent with the fact that there is no dependence on 
$\alpha_3, \phi$ and $\psi$ in the Ricci tensor components in the left
hand side of equation.

%%%%%%%%%%%%%%%%%%%%%%%%%%%%%%%%%%%%%%%%%%%%%%%%%%%%%%%%%%%%%%%%%%%%%%%%%%

\end{document}